\documentclass[runningheads,smallsec]{llncs}

\usepackage{underscore}
\usepackage[ruled,lined,vlined]{algorithm2e}
\usepackage{algpseudocode}
\usepackage{pifont}
\usepackage{lineno}
\usepackage{cite}
\usepackage{paralist}
\usepackage{mathtools}
\usepackage{xspace}
\usepackage{amssymb}
\usepackage{amsmath}
\usepackage{amsthm}
\usepackage{thmtools}
\usepackage{thm-restate}
\usepackage{wrapfig}
\usepackage{bm}
\usepackage[bookmarks,unicode,colorlinks=true]{hyperref}%
   \def\@citecolor{blue}%
   \def\@urlcolor{blue}%
   \def\@linkcolor{blue}%
\def\orcidID#1{\smash{\href{http://orcid.org/#1}{\protect\raisebox{-1.25pt}{\protect\includegraphics{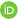}}}}}
\usepackage[dvipsnames]{xcolor}

\iftrue
	\usepackage{todonotes}
	\newcommand{\sven}[1]{\todo[inline,color=teal!10,caption={Sven}]{\textbf{Sven:} #1}}
    \newcommand{\ly}[1]{\todo[inline,color=orange!10,caption={LY}]{\textbf{LY:} #1}}
	\newcommand{\myv}[1]{\todo[inline,color=green!10,caption={MYV}]{\textbf{MYV:} #1}}
 	\newcommand{\qiyi}[1]{\todo[inline,color=magenta!10,caption={Qiyi}]{\textbf{Qiyi:} #1}}
\else
	\newcommand{\sven}[1]{}
	\newcommand{\ly}[1]{}
	\newcommand{\myv}[1]{}
 	\newcommand{\qiyi}[1]{}

\fi

\newcommand{\A}{\mathcal{A}}
\newcommand{\B}{\mathcal{B}}
\newcommand{\C}{\mathcal{C}}
\newcommand{\sd}{\mathcal{S}}
\newcommand{\D}{\mathcal{D}}

\newcommand{\F}{\mathcal{F}}
\newcommand{\R}{\mathcal{R}}
\newcommand{\M}{\mathcal{M}}
\newcommand{\N}{\mathcal{N}}

\newcommand{\T}{\mathcal{T}}

\newcommand{\G}{\mathcal{G}}
\newcommand{\bigO}{\mathcal{O}}

\newcommand{\wordletter}[2]{#1{[#2]}}
\newcommand{\subword}[3]{#1 {[#2..#3]}}

\newcommand{\pathto}[2]{{\xrightarrow[]{{#1}}}}

\newcommand{\alphabet}{\Sigma}
\newcommand{\emptyword}{\epsilon}

\newcommand{\finwords}{\alphabet^*}
\newcommand{\infwords}{\alphabet^\omega}
\newcommand{\poswords}{\alphabet^+}
\newcommand{\langsymb}[0]{\mathcal{L}}
\newcommand{\lang}[1]{\langsymb(#1)}
\newcommand{\finlang}[1]{\langsymb_{*}(#1)}
\newcommand{\inflang}[1]{\langsymb(#1)}

\newcommand{\upword}[1]{\text{UP}(#1)}

\newcommand{\states}{Q}

\newcommand{\trans}{\delta}
\newcommand{\Trans}{\Delta}

\newcommand{\init}{q_0}

\newcommand{\accd}{\Gamma}

\newcommand{\run}{\rho}

\newcommand{\naturals}{\mathbb{N}}
\newcommand{\ndet}{\mathsf{ndet}}

\newcommand{\setnocond}[1]{\{#1\}}
\newcommand{\setcond}[2]{\{\, #1 \mid #2 \,\}}

\newcommand{\infof}{\mathsf{inf}}

\newcommand{\buchi}{B\"uchi\xspace}
\renewcommand{\@}{\xspace}

\newcommand{\size}[1]{|#1|}
\newcommand{\canoEq}{\backsim}

\newcommand{\proEq}{\approx}

\newcommand{\proclassv}[2]{[#1]_{{\proEq}_{#2}}}
\newcommand{\proclassc}[2]{[#1]_{{\proEq}^C_{#2}}}

\newcommand{\class}[1]{[#1]_{\canoEq}}

\newcommand{\quotient}{\finwords/_{\canoEq}}

\newcommand{\fn}[1]{\texttt{#1}}
\newcommand{\tfn}[1]{\textit{#1}}
\newcommand{\rep}[1]{\Tilde{#1}}

\title{A novel family of finite automata for recognizing and learning $\omega$-regular languages}
\author{Yong Li\orcidID{0000-0002-7301-9234}, Sven Schewe\orcidID{0000-0002-9093-9518}, Qiyi Tang\orcidID{0000-0002-9265-3011}}
\institute{University of Liverpool, UK}
\titlerunning{Novel Families of Finite Automata}

\begin{document}\sloppy
\maketitle

\begin{abstract}
    Families of DFAs (FDFAs) have recently been introduced as a new representation of $\omega$-regular languages.
    They target ultimately periodic words, with acceptors revolving around accepting some representation $u\cdot v^\omega$.
    Three canonical FDFAs have been suggested, called \emph{periodic}, \emph{syntactic}, and \emph{recurrent}.
    We propose a fourth one, \emph{limit FDFAs}, which can be exponentially coarser than periodic FDFAs and are more succinct than syntactic FDFAs, while they are incomparable (and dual to) recurrent FDFAs.
    We show that limit FDFAs can be easily used to check not only whether $\omega$-languages are regular, but also whether they are accepted by deterministic \buchi automata.
    We also show that canonical forms can be left behind in applications: the limit and recurrent FDFAs can complement each other nicely, and it may be a good way forward to use a combination of both.
    Using this observation as a starting point, we explore making more efficient use of Myhill-Nerode's right congruences in aggressively increasing the number of don't-care cases in order to obtain smaller progress automata. In pursuit of this goal, we gain succinctness, but pay a high price by losing constructiveness.
    
\end{abstract}

\section{Introduction}
The class of $\omega$-regular languages has proven to be an important formalism to model reactive systems and their specifications, and automata over infinite words are the main tool to reason about them.
For example, the automata-theoretic approach to verification~\cite{DBLP:conf/lics/VardiW86} is the main framework for verifying $\omega$-regular specifications.
The first type of automata recognizing $\omega$-regular languages is nondeterministic \buchi automata~\cite{Buc62} (NBAs) where an infinite word is accepted if one of its runs meets the accepting condition for infinitely many times.
Since then, other types of acceptance conditions, such as Muller, Rabin, Streett and parity automata~\cite{DBLP:books/ems/21/WilkeS21}, have been introduced.
All the automata mentioned above are finite automata processing \emph{infinite} words, widely known as $\omega$-automata~\cite{DBLP:books/ems/21/WilkeS21}.

The theory of $\omega$-regular languages is more involved than that of regular languages.
For instance, nondeterministic finite automata (NFAs) can be determinized with a subset construction, while NBAs have to make use of tree structures~\cite{DBLP:conf/focs/Safra88}.
This is because of a fundamental difference between these language classes: for a given regular language $R$, the Myhill-Nerode theorem~\cite{Myhill57,Nerode58} defines a right congruence (RC) $\canoEq_R$ in which every equivalence class corresponds to a state in the minimal deterministic finite automata (DFA) accepting $R$.
In contrast, there is no similar theorem to define the minimal deterministic $\omega$-automata for the full class of $\omega$-regular languages\footnote{Simple extension of Myhill-Nerode theorem for $\omega$-regular languages only works on a small subset~\cite{DBLP:journals/iandc/MalerP95,DBLP:journals/iandc/AngluinF21}}.
Schewe proved in~\cite{DBLP:conf/fsttcs/Schewe10} that it is NP-complete to find the minimal deterministic $\omega$-automaton even given a deterministic $\omega$-automaton.
Therefore, it seems impossible to easily define a Myhill-Nerode theorem for (minimal) $\omega$-automata.

Recently, Angluin, Boker and Fisman~\cite{AngluinBF18} proposed families of DFAs (FDFAs) for recognizing $\omega$-regular languages, in which every DFA can be defined with respect to a RC defined over a given $\omega$-regular language~\cite{DBLP:journals/tcs/AngluinF16}.
This tight connection is the theoretical foundation on which the state of the art learning algorithms for $\omega$-regular languages~\cite{DBLP:journals/tcs/AngluinF16,DBLP:journals/iandc/LiCZL21} using membership and equivalence queries~\cite{DBLP:journals/iandc/Angluin87} are built. 
FDFAs are based on well-known properties of $\omega$-regular languages~\cite{Buc62,DBLP:conf/mfps/CalbrixNP93}:
two $\omega$-regular languages are equivalent if, and only if, they have the same set of \emph{ultimately periodic words}.
An ultimately periodic word $w$ is an infinite word that consists of first a finite prefix $u$, followed by an infinite repetition of a finite nonempty word $v$; it can thus be represented as a decomposition pair $(u, v)$. 
FDFAs accept infinite words by accepting their decomposition pairs:
an FDFA $\F = (\M, \setnocond{\N^{q}})$ consists of a \emph{leading DFA} $\M$ that processes the finite prefix $u$, while leaving the acceptance work of $v$ to the \emph{progress DFA} $\N^q$, one for each state of $\M$.
To this end, $\M$ intuitively tracks the Myhill-Nerode's RCs, and an ultimately periodic word $u\cdot v^\omega$ is accepted if it has a representation $x \cdot y^\omega$ such that $x$ and $x \cdot y$ are in the same congruence class and $y$ is accepted by the progress DFA $\N^x$.
Angluin and Fisman~\cite{DBLP:journals/tcs/AngluinF16} formalized the RCs of three canonical FDFAs, namely periodic~\cite{DBLP:conf/mfps/CalbrixNP93}, syntactic~\cite{DBLP:journals/tcs/MalerS97} and recurrent~\cite{DBLP:journals/tcs/AngluinF16}, and provided a unified learning framework for them.

In this work, we first propose a fourth one, called \emph{limit FDFAs} (cf. Section~\ref{sec:limit-fdfas-def}).
We show that limit FDFAs are coarser than syntactic FDFAs.
Since syntactic FDFAs can be exponentially more succinct than periodic FDFAs~\cite{DBLP:journals/tcs/AngluinF16}, so do our limit FDFAs.
We show that limit FDFAs are dual (and thus incomparable in the size) to recurrent FDFAs, due to symmetric treatment for don't care words.
More precisely, the formalization of such FDFA does not care whether or not a progress automaton $\N^x$ accepts or rejects a word $v$, unless reading it in $\M$ produces a self-loop.  
Recurrent progress DFAs reject all those don't care words, while limit progress DFAs accept them.

We show that limit FDFAs (families of DFAs that use limit DFAs) have two interesting properties.
The first is on conciseness:
we show that this change in the treatment of don't care words not only defines a dual to recurrent FDFAs but also allows us to identify languages accepted by deterministic \buchi automata (DBAs) easily.
It is only known that one can identify whether a given $\omega$-language is regular by verifying whether the number of states in the three canonical FDFAs is finite.
However, if one wishes to identify DBA-recognizable languages with FDFAs, a straight-forward approach is to first translate the input FDFA to an equivalent deterministic Rabin automaton~\cite{AngluinBF18} through an intermediate NBA, and then use the deciding algorithm in \cite{DBLP:conf/isaac/KrishnanPB94} by checking the transition structure of Rabin automata.
However, this approach is exponential in the size of the input FDFA because of the NBA determinization procedure~\cite{DBLP:conf/focs/Safra88,DBLP:conf/fossacs/Schewe09,DBLP:conf/icalp/ColcombetZ09}.
Our limit FDFAs are, to the best of our knowledge, the \emph{first} type of FDFAs able to identify the DBA-recognizable languages in polynomial time (cf. Section~\ref{sec:dba-recognizable}).

We note that limit FDFAs also fit nicely into the learning framework introduced in \cite{DBLP:journals/tcs/AngluinF16}, so that they can be used for learning without extra development.

We then discuss how to make more use of don't care words when defining the RCs of the progress automata, leading to the coarsest congruence relations and therefore the most concise FDFAs, albeit to the expense of losing constructiveness (cf. Section~\ref{sec:progress-RP}).

\section{Preliminaries}

In the whole paper, we fix a finite \emph{alphabet} $\alphabet$.
A \emph{word} is a finite or infinite sequence of letters in $\alphabet$;
$\emptyword$ denotes the empty word.
Let $\finwords$ and $\infwords$ denote the set of all finite and infinite words (or $\omega$-words), respectively.
In particular, we let $\poswords = \finwords\setminus\setnocond{\emptyword}$.
A \emph{finitary language} is a subset of $\finwords$; 
an \emph{$\omega$-language} is a subset of $\infwords$.
Let $\run$ be a sequence; 
we denote by $\wordletter{\run}{i}$ the $i$-th element of $\run$ and by $\subword{\run}{i}{k}$ the subsequence of $\run$ starting at the $i$-th element and ending at the $k$-th element (inclusively) when $i \leq k$, and the empty sequence $\emptyword$ when $i > k$.
Given a finite word $u$ and a word $w$, we denote by $u \cdot w$ ($uw$, for short) the concatenation of $u$ and $w$. 
Given a finitary language $L_{1}$ and a finitary/$\omega$-language $L_{2}$, the concatenation $L_{1} \cdot L_{2} $ ($L_{1} L_{2}$, for short) of $L_{1}$ and $L_{2}$ is the set $L_{1} \cdot L_{2} = \setcond{uw}{u \in L_{1}, w \in L_{2}}$ and $L^{\omega}_{1}$ the infinite concatenation of $L_{1}$.

\paragraph{\bf Transition system.}
A (nondeterministic) transition system (TS) is a tuple $\T = (\states, \init, \trans)$, where $\states$ is a finite set of states, $\init \in \states$ is the initial state, and $\trans: \states \times \alphabet \rightarrow 2^{\states}$ is a transition function. 
We also lift $\trans$ to sets as 
$\trans (S, \sigma) := \bigcup_{q\in S}\trans(q, \sigma)$.
We also extend $\trans$ to words, by letting $\trans(S, \emptyword) = S$ and $\trans(S, a_{0} a_{1} \cdots a_{k}) = \trans(\trans(S, a_{0}), a_1 \cdots\@  a_{k})$, where we have $k \geq 1$ and $a_{i} \in \alphabet$ for $i \in \setnocond{0, \cdots\@ , k}$.

The \emph{underlying graph} $\G_{\T}$ of a TS $\T$ is a graph $\langle \states, E\rangle$, where the set of vertices is the set $\states$ of states in $\T$ and $(q, q') \in E$ if $q' \in \trans(q, a)$ for some $a \in \alphabet$.
We call a set $C \subseteq \states$ a \emph{strongly connected component} (SCC) of $\T$ if, for every pair of states $q, q' \in C$, $q$ and $q'$  can reach each other in $\G_{\T}$.

\paragraph{\bf Automata.}
An automaton on finite words is called a \emph{nondeterministic finite automaton} (NFA).
An NFA $\A$ is formally defined as a tuple $(\T, F)$, where $\T$ is a TS and $F\subseteq \states$ is a set of \emph{final} states.
An automaton on $\omega$-words is called a \emph{nondeterministic \buchi automaton} (NBA).
An NBA $\B$ is represented as a tuple $(\T, \accd)$ where $\T$ is a TS and $\accd \subseteq \setnocond{(q, a, q'): q,q'\in \states, a\in \alphabet, q'\in\trans(q,a)}$ is a set of \emph{accepting} transitions.
An NFA $\A$ is said to be a \emph{deterministic} finite automaton (DFA) if, for each $q \in \states$ and $a \in \alphabet$, $\size{\trans(q, a)} \leq 1$. 
Deterministic \buchi automata (DBAs) are defined similarly and thus $\accd $ is a subset of $ \setnocond{(q, a): q\in\states, a \in \alphabet}$, since the successor $q'$ is determined by the source state and the input letter. 

A \emph{run} of an NFA $\A$ on a finite word $u$ of length $n \geq 0$ is a sequence of states $\run = q_{0} q_{1} \cdots q_{n} \in \states^{+}$ such that, for every $0 \leq i < n$, $q_{i+1} \in \trans(q_{i}, \wordletter{u}{i})$.
We write $q_{0} \pathto{u}{\trans} q_{n}$ if there is a run from $q_{0}$ to $q_{n}$ over $u$.
A finite word $u \in \finwords$ is \emph{accepted} by an NFA $\A$ if there is a run $q_{0} \cdots q_{n}$ over $u$ such that $q_{n} \in F$. 
Similarly, an \emph{$\omega$-run} of $\A$ on an $\omega$-word $w$ is an infinite sequence of transitions $\run = (q_{0}, \wordletter{w}{0}, q_{1}) (q_1, \wordletter{w}{1}, q_2)\cdots$ such that, for every $i \geq 0$, $q_{i+1} \in \trans(q_{i}, w[i])$. 
Let $\inf(\run)$ be the set of transitions that occur infinitely often in the run $\run$.
An $\omega$-word $w \in \infwords$ is \emph{accepted} by an NBA $\A$ if there exists an $\omega$-run $\run$ of $\A$ over $w$ such that $\inf({\run}) \cap \accd \neq \emptyset$. 
The \emph{finitary language} recognized by an NFA $\A$, denoted by $\finlang{\A}$, is defined as the set of finite words accepted by it.
Similarly, we denote by $\inflang{\A}$ the \emph{$\omega$-language} recognized by an NBA $\A$, i.e., the set of $\omega$-words accepted by $\A$.
NFAs/DFAs accept exactly \emph{regular} languages while NBAs recognize exactly \emph{$\omega$-regular} languages.

\paragraph{\bf Right congruences.}
A \emph{right congruence} (RC) relation is an equivalence relation $\canoEq$ over $\finwords$ such that $x \canoEq y$ implies $xv \canoEq yv$ for all $v \in \finwords$.
We denote by $\size{\canoEq}$ the index of $\canoEq$, i.e.\@, the number of equivalence classes of $\canoEq$.
A \emph{finite RC} is a RC with a finite index.
We denote by $\quotient$ the set of equivalence classes of $\finwords$ under $\canoEq$.
Given $x \in \finwords$, we denote by $\class{x}$ the equivalence class of $\canoEq$ that $x$ belongs to.

For a given RC $\canoEq$ of a regular language $R$, the Myhill-Nerode theorem~\cite{Myhill57,Nerode58} defines a unique minimal DFA $D$ of $R$, in which each state of $D$ corresponds to an equivalence class defined by $\canoEq$ over $\finwords$.
Therefore, we can construct a DFA $\D[\canoEq]$ from $\canoEq$ in a standard way.
\begin{definition}[\hspace*{-1.5mm}\cite{Myhill57,Nerode58}]
\label{def:induced-dfw}
    Let $\canoEq$ be a right congruence of finite index.
    The TS $\T[\canoEq]$ induced by $\canoEq$ is a tuple $(S, s_{0}, \trans)$ where $S = \quotient$, $s_{0} = \class{\emptyword}$, and for each $u \in \finwords$ and $a \in \alphabet$, $\trans(\class{u}, a) = \class{ua}$.
\end{definition}

For a given regular language $R$, we can define the RC $\canoEq_{R}$ of $R$ as
$x \canoEq_{R} y \text{ if, and only if, } \forall v \in \finwords.\ xv \in R \Longleftrightarrow yv \in R$. 
Therefore, the minimal DFA for $R$ is the DFA $\D[\canoEq_{R}] = (\T[\canoEq_R], F_{\canoEq_R})$ by setting final states $F_{\canoEq_R}$ to all equivalence classes $[u]_{\canoEq_R}$ such that $u \in R$.

\paragraph{\bf Ultimately periodic (UP) words.}
A UP-word $w$ is an $\omega$-word of the form $uv^{\omega}$, where $u \in \finwords$ and $v \in \poswords$.
Thus $w = uv^{\omega}$ can be represented as a pair of finite words $(u, v)$, called a \emph{decomposition} of $w$.
A UP-word can have multiple decompositions: 
for instance $(u, v)$, $(uv, v)$, and $(u, vv)$ are all decompositions of $uv^{\omega}$.
For an $\omega$-language $L$, let $\upword{L} = \setcond{uv^{\omega} \in L}{u \in \finwords \land v \in \poswords}$ denote the set of all UP-words in $L$.
The set of UP-words of an $\omega$-regular language $L$ can be seen as the fingerprint of $L$, as stated below.
\begin{theorem}[\hspace*{-1.3mm}\cite{Buc62,DBLP:conf/mfps/CalbrixNP93}]
\label{thm:upword-omega-regular-language}
    (1)
    Every non-empty $\omega$-regular language $L$ contains at least one UP-word.
    (2)		
    Let $L$ and $L'$ be two $\omega$-regular languages.
    Then $L = L'$ if, and only if, $\upword{L} = \upword{L'}$.
\end{theorem}

\paragraph{\bf Families of DFAs (FDFAs).}
Based on Theorem~\ref{thm:upword-omega-regular-language}, Angluin, Boker, and Fisman~\cite{AngluinBF18} introduced the notion of FDFAs to recognize $\omega$-regular languages.
\begin{definition}[\hspace*{-1.3mm}\cite{AngluinBF18}]
\label{def:fdfws}
    An FDFA is a pair $\F = (\M, \setnocond{\N^{q}})$ consisting of a leading DFA $\M$ and of a progress DFA $\N^{q}$ for each state $q$ in $\M$.
\end{definition}

Intuitively, the leading DFA $\M$ of $\F = (\M, \setnocond{\N^{q}})$ for $L$ consumes the finite prefix $u$ of a UP-word $uv^{\omega} \in \upword{L}$, reaching some state $q$ and, for each state $q$ of $\M$, the progress DFA $\N^{q}$ accepts the period $v$ of $uv^{\omega}$.
Note that the leading DFA $\M$ of every FDFA does not make use of final states---contrary to its name, it is really a leading transition system.

Let $A$ be a deterministic automaton with TS $\T = (\states, q_0, \trans)$ and $x \in \finwords$.
We denote by $A(x)$ the state $\trans(q_0, x)$.
Each FDFA $\F$ characterizes a set of UP-words $\upword{\F}$ by following the acceptance condition.
\begin{definition}[Acceptance]
\label{def:acc-fdfa}
    Let $\F = (\M, \setnocond{\N^{q}})$ be an FDFA and $w$ be a UP-word.
    A decomposition $(u, v)$ of $w$ is \emph{normalized} with respect to $\F$ if $\M(u) = \M(uv)$.
    A decomposition $(u, v)$ is accepted by $\F$ if $(u, v)$ is normalized and we have $v \in \finlang{\N^{q}}$ where $q = \M(u)$.
    The UP-word $w$ is accepted by $\F$ if there exists a decomposition $(u, v)$ of $w$ accepted by $\F$.
\end{definition}

Note that the acceptance condition in~\cite{AngluinBF18} is defined with respect to the decompositions, while ours applies to UP-words.
So, they require the FDFAs to be saturated for recognizing $\omega$-regular languages.
\begin{definition}[Saturation~\cite{AngluinBF18}]
\label{def:saturated-fdfws}
    Let $\F$ be an FDFA and $w$ be a UP-word in $\upword{\F}$.
    We say $\F$ is \emph{saturated} if, for all normalized decompositions $(u, v)$ and $(u', v')$ of $w$, either both $(u, v)$ and $(u', v')$ are accepted by $\F$, or both are not.
\end{definition}
We will see in Section~\ref{ssec:dba-limit-fdfa} that under our acceptance definition the saturation property can be relaxed while still accepting the same language.

In the remainder of the paper, we 
fix an $\omega$-language $L$ unless stated otherwise.
\section{Limit FDFAs for recognizing $\omega$-regular languages}
\label{sec:limit-fdfas-def}

In this section, we will first recall the definitions of three existing canonical FDFAs for $\omega$-regular languages, and then introduce our limit FDFAs and compare the four types of FDFAs.

\subsection{Limit FDFAs and other canonical FDFAs}
Recall that, for a given regular language $R$, by Definition~\ref{def:induced-dfw}, the Myhill-Nerode theorem~\cite{Myhill57,Nerode58} associates each equivalence class of $\canoEq_R$ with a state of the minimal DFA $\D[\canoEq_R]$ of $R$.
The situation in $\omega$-regular languages is, however, more involved~\cite{DBLP:journals/iandc/AngluinF21}.
An immediate extension of such RCs for an $\omega$-regular language $L$ is the following.
\begin{definition}[Leading RC]\label{def:leading-rc}
    For two $u_1, u_2 \in \finwords$, $u_1 \canoEq_L u_2$ if, and only if $\forall w \in \infwords$. $u_1 w \in L \Longleftrightarrow u_2 w \in L$. 
\end{definition}

Since we fix an $\omega$-language $L$ in the whole paper, we will omit the subscript in $\canoEq_L$ and directly use $\canoEq$ in the remainder of the paper.

Assume that $L$ is an $\omega$-regular language.
Obviously, the index of $\canoEq$ is \emph{finite} since it is not larger than the number of states in the minimal deterministic $\omega$-automaton accepting $L$. 
However, $\canoEq$ is only enough to define the minimal $\omega$-automaton for a small subset of $\omega$-regular languages; see~\cite{DBLP:journals/iandc/MalerP95,DBLP:journals/iandc/AngluinF21} for details about such classes of languages.
For instance, consider the language $L = (\finwords \cdot aa)^{\omega}$ over $\alphabet = \setnocond{a, b}$:
clearly, $|\canoEq| = 1$ because $L$ is a suffix language (for all $u \in \finwords$, $w \in L \Longleftrightarrow u \cdot w \in L$).
At the same time, it is easy to see that the minimal deterministic $\omega$-automaton needs at least two states to recognize $L$.
Hence, $\canoEq$ alone does not suffice to recognize the full class of $\omega$-regular languages.

Nonetheless, based on Theorem~\ref{thm:upword-omega-regular-language}, we only need to consider the UP-words when uniquely identifying a given $\omega$-regular language $L$ with RCs.
Calbrix \emph{et al.} proposed in~\cite{DBLP:conf/mfps/CalbrixNP93} the use of the regular language $L_{\$} = \setnocond{ u \$ v: u \in \finwords, v \in \poswords, uv^{\omega} \in L}$ to represent $L$, where $\$ \notin \alphabet$ is a fresh letter\footnote{This enables to learn $L$ via learning the regular language $L_{\$}$~\cite{DBLP:conf/tacas/FarzanCCTW08}.}.
Intuitively, $L_{\$}$ associates a UP-word $w$ in $\upword{L}$ by containing every decomposition $(u, v)$ of $w$ in the form of $u\$ v$.
The FDFA representing $L_{\$}$ is formally stated as below.
\begin{definition}[Periodic FDFAs~\cite{DBLP:conf/mfps/CalbrixNP93}]\label{def:periodic-fdfa}
The $\canoEq$ is as defined in Definition~\ref{def:leading-rc}.

Let $[u]_{\canoEq}$ be an equivalence class of $\canoEq$.
For $x, y \in \finwords$, we define periodic RC as: $ x \proEq^{u}_{P} y$ if, and only if, $\forall v \in \finwords$, $u\cdot (x \cdot v)^{\omega} \in L \Longleftrightarrow u\cdot (y\cdot v)^{\omega} \in L$.  

The periodic FDFA $\F_P = (\M, \setnocond{\N^{u}_P})$ of $L$ is defined as follows.

The leading DFA $\M$ is the tuple $(\T[\canoEq], \emptyset)$. Recall that $\T[\canoEq]$ is the TS constructed from $\canoEq$ by Definition~\ref{def:induced-dfw}.

The periodic progress DFA $\N^{u}_P$ of the state $[u]_{\canoEq} \in \finwords/_{\canoEq}$ is the tuple $(\T[\proEq^u_P], F_u)$, where $[v]_{\proEq^u_P} \in F_u$ if $uv^{\omega} \in L$.
\end{definition}
One can verify that, for all $u, x, y, v \in \finwords$, if $x \proEq^u_P y$, then $xv \proEq^u_P y v$.
Hence, $\proEq^u_P$ is a RC.
It is also proved in~\cite{DBLP:conf/mfps/CalbrixNP93} that $L_{\$}$ is a regular language, so the index of $\proEq^u_P$ is also finite. 

Angluin and Fisman in \cite{DBLP:journals/tcs/AngluinF16} showed that, for a variant of the family of languages $L_n$ given by Michel~\cite{michel1988complementation}, its periodic FDFA has $\Omega(n!)$ states, while the syntactic FDFA obtained in \cite{DBLP:journals/tcs/MalerS97} only has $\bigO(n^2)$ states.
 The leading DFA of the syntactic FDFAs is exactly the one defined for the periodic FDFA.
 The two types of FDFAs differ in the definitions of the progress
DFAs $\N^u$ for some $[u]_{\canoEq}$.
From Definition~\ref{def:periodic-fdfa}, one can see that $\N^u_P$ accepts the finite words in $V_u = \setnocond{v \in \poswords: u \cdot v^{\omega} \in L}$.
The progress DFA $\N^u_S$ of the syntactic FDFA is not required to accept all words in $V_u$, but only a subset $V_{u,v} = \setnocond{v \in \poswords: u\cdot v^{\omega} \in L, u \canoEq u \cdot v}$, over which the leading DFA $\M$ can take a round trip from $\M(u)$ back to itself.
This minor change makes the syntactic FDFAs of the language family $L_n$ exponentially more succinct than their periodic counterparts.

Formally, syntactic FDFAs are defined as follows.

\begin{definition}[Syntactic FDFA~\cite{DBLP:journals/tcs/MalerS97}]\label{def:syntactic-rcs}
The $\canoEq$ is as defined in Definition~\ref{def:leading-rc}.

Let $[u]_{\canoEq}$ be an equivalence class of $\canoEq$.
For $x, y \in \finwords$, we define syntactic RC as: $ x \proEq^u_{S} y$ if and only if  $u \cdot x \canoEq u \cdot y$ and for $\forall v \in \finwords$, if $u\cdot x \cdot v \canoEq u $, then $ u\cdot (x \cdot v)^{\omega} \in L \Longleftrightarrow u\cdot (y\cdot v)^{\omega} \in L$.  

The syntactic FDFA $\F_S = (\M, \setnocond{\N^{u}_S})$ of $L$ is defined as follows.

The leading DFA $\M$ is the tuple $(\T[\canoEq], \emptyset)$ as defined in Definition~\ref{def:periodic-fdfa}.

The syntactic progress DFA $\N^{u}_S$ of the state $[u]_{\canoEq} \in \finwords/_{\canoEq}$ is the tuple $(\T[\proEq^u_S], F_u)$ where $[v]_{\proEq^u_S} \in F_u$ if $u \cdot v \canoEq u$ and $uv^{\omega} \in L$.
\end{definition}

Angluin and Fisman \cite{DBLP:journals/tcs/AngluinF16} noticed that the syntactic progress RCs are not defined with respect to the regular language $V_{u,v} = \setnocond{v \in \poswords: u\cdot v^{\omega} \in L, u \canoEq u \cdot v}$ as $\canoEq_{V_{u,v}}$ that is similar to $\canoEq_R$ for a regular language $R$. They proposed the recurrent progress RC $\proEq^u_R$ that mimics the RC $\canoEq_{V_{u,v}}$ to obtain a DFA accepting $V_{u, v}$ as follows.
\begin{definition}[Recurrent FDFAs~\cite{DBLP:journals/tcs/AngluinF16}]
The $\canoEq$ is as defined in Definition~\ref{def:leading-rc}.

Let $[u]_{\canoEq}$ be an equivalence class of $\canoEq$.
For $x, y \in \finwords$, we define recurrent RC as: $ x \proEq^u_{R} y$ if and only if $\forall v \in \finwords$, $(u \cdot x \cdot v \canoEq u \land u\cdot (x v)^{\omega} \in L) \Longleftrightarrow (u\cdot y v \canoEq u \land u\cdot (y \cdot v)^{\omega} \in L)$.  

The recurrent FDFA $\F_R = (\M, \setnocond{\N^{u}_R})$ of $L$ is defined as follows.

The leading DFA $\M$ is the tuple $(\T[\canoEq], \emptyset)$ as defined in Definition~\ref{def:periodic-fdfa}.

The recurrent progress DFA $\N^{u}_R$ of the state $[u]_{\canoEq} \in \finwords/_{\canoEq}$ is the tuple $(\T[\proEq^u_R], F_u)$ where $[v]_{\proEq^u_R} \in F_u$ if $u \cdot v \canoEq u$ and $uv^{\omega} \in L$.
\end{definition}

As pointed out in \cite{DBLP:journals/tcs/AngluinF16}, the recurrent FDFAs may \emph{not} be minimal because, according to Definition \ref{def:acc-fdfa}, FDFAs only care about the normalized decompositions, i.e, whether a word in $C_u = \setnocond{v \in \poswords: u \cdot v \canoEq u}$ is accepted by the progress DFA $\N^u_R$.
However, there are \emph{don't care} words that are not in $C_u$ and recurrent FDFAs treat them all as \emph{rejecting}%
\footnote{Minimizing DFAs with don't care words is NP-complete~\cite{DBLP:journals/tc/Pfleeger73}}.

Our argument is that the don't care words are \emph{not} necessarily rejecting and can also be regarded as \emph{accepting}.
This idea allows the progress DFAs $\N^u$ to accept the regular language $\setnocond{v \in \poswords: u \cdot v \canoEq u \implies u \cdot v^{\omega} \in L}$, rather than $\setnocond{v \in \poswords: u \cdot v \canoEq u \land u \cdot v^{\omega} \in L}$.
This change allows a translation of limit FDFAs to DBAs with a quadratic blow-up when $L$ is DBA-recognizable language, as shown later in Section \ref{sec:dba-recognizable}.
We formalize this idea as below and define a new type of FDFAs called \emph{limit FDFAs}.

\begin{definition}[Limit FDFAs]\label{def:limit-rcs}
The $\canoEq$ is as defined in Definition~\ref{def:leading-rc}.

Let $[u]_{\canoEq}$ be an equivalence class of $\canoEq$.
For $x,y \in \finwords$, we define limit RC as: $ x \proEq^{u}_L y$ if and only if $\forall v \in \finwords$, $(u \cdot x \cdot v \canoEq u \Longrightarrow u\cdot (x \cdot v)^{\omega} \in L) \Longleftrightarrow (u \cdot y\cdot v \canoEq u \Longrightarrow u\cdot (y \cdot v)^{\omega} \in L)$.  

The limit FDFA $\F_L = (\M, \setnocond{\N^{u}_L})$ of $L$ is defined as follows.

The leading DFA $\M$ is the tuple $(\T[\canoEq], \emptyset)$ as defined in Definition~\ref{def:periodic-fdfa}.

The progress DFA $\N^{u}_L$ of the state $[u]_{\canoEq} \in \finwords/_{\canoEq}$ is the tuple $(\T[\proEq^u_L], F_u)$ where $[v]_{\proEq^u_L} \in F_u$ if $u \cdot v \canoEq u \implies uv^{\omega} \in L$.
\end{definition}

We need to show that $\proEq^u_L$ is a RC.
For $u, x, y , v' \in \finwords$, if $x \proEq^u_L y$, we need to prove that $xv' \proEq^u_L yv'$, i.e., for all $e \in \finwords$, $(u \cdot xv' \cdot e \canoEq u \implies u \cdot (xv'\cdot e)^{\omega} \in L) \Longleftrightarrow (u \cdot yv' \cdot e \canoEq u \implies u \cdot (yv'\cdot e)^{\omega} \in L)$.
This follows immediately from the fact that $x \proEq^u_L y$ by setting $v = v'\cdot e$ for all $e\in \finwords$ in Definition \ref{def:limit-rcs}.

\begin{figure}[t]
    \centering
    \includegraphics[scale=0.70]{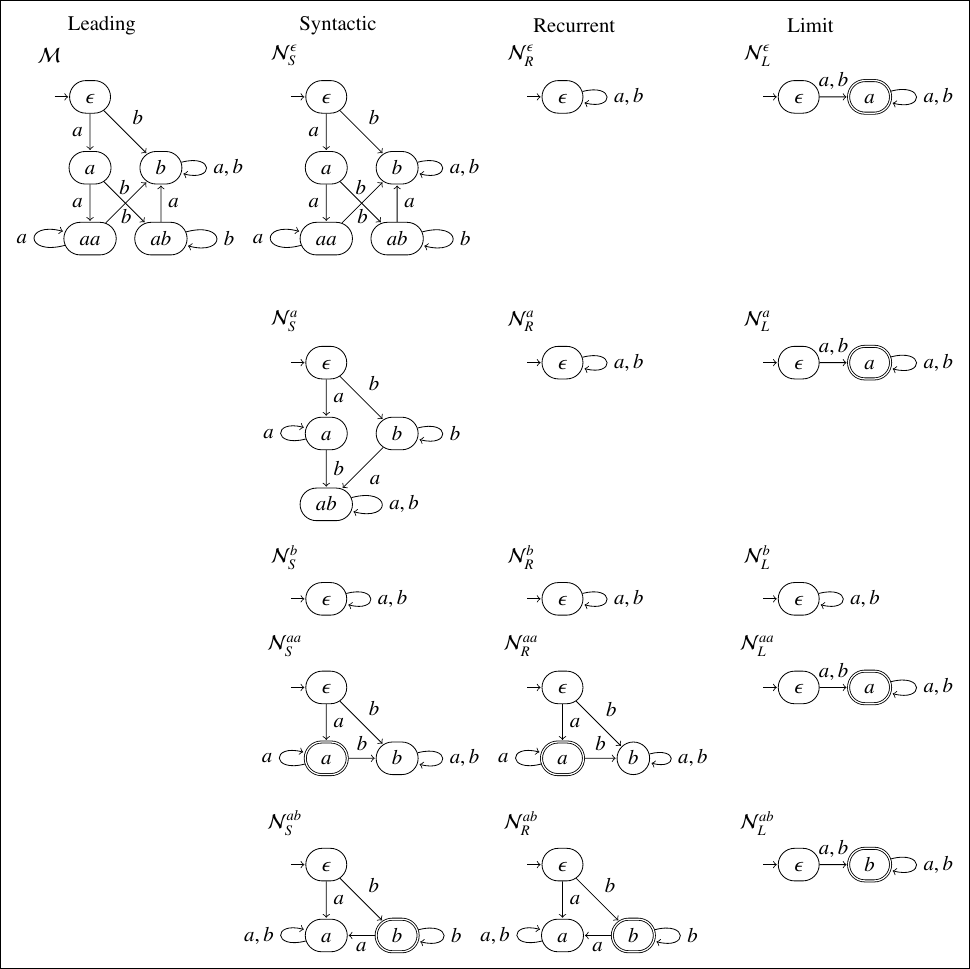}
    \caption{Three types of FDFAs for $L = a^{\omega} + ab^{\omega}$. The final states are marked with double lines.}
    \label{fig:limit-fdfa-example}
\end{figure}

Let $L = a^{\omega} + ab^{\omega}$ be a language over $\alphabet = \setnocond{a, b}$.
Three types of FDFAs are depicted in Figure \ref{fig:limit-fdfa-example}, where the leading
DFA $\M$ is given in the column labeled with ''Leading” and the progress DFAs are in the column
labeled with ``Syntactic", ``Recurrent" and ``Limit".
We omit the periodic FDFA here since we will focus more on the other three in this work.
Consider the progress DFA $\N^{aa}_L$:
there are only two equivalence classes, namely $[\emptyword]_{\proEq^{aa}_L}$ and $[a]_{\proEq^{aa}_L}$.
We can use $v = \emptyword$ to distinguish $\emptyword$ and a word $x \in \poswords$ since $aa \cdot \emptyword \canoEq aa \implies aa \cdot (\emptyword \cdot \emptyword)^{\omega} \in L$ does not hold, while $aa \cdot x \canoEq aa \implies aa \cdot (x \cdot \emptyword)^{\omega} \in L$ holds.
For all $x, y \in \poswords$, $x \proEq^{aa}_L y$ since both $aa \cdot x \canoEq aa \implies aa \cdot (x \cdot v)^{\omega} \in L$ and $aa \cdot y \canoEq aa \implies aa \cdot (y\cdot v)^{\omega} \in L$ hold for all $v \in \finwords$.
One can also verify the constructions for the syntactic and recurrent progress DFAs.
We can see that the don't care word $b$ for the class $[aa]_{\canoEq}$ are rejecting in both $\N_S^{aa}$ and $\N^{aa}_R$, while it is accepted by $\N^{aa}_L$.
Even though $b$ is accepted in $\N^{aa}_L$, one can observe that $(aa, b)$ (and thus $aa\cdot b^{\omega}$) is not accepted by the limit FDFA, according to Definition~\ref{def:acc-fdfa}.
Indeed, the three types of FDFAs still recognize the same language $L$.

When the index of $\canoEq$ is only one, then $\emptyword \canoEq u$ holds for all $u \in \finwords$.
Corollary~\ref{coro:trivial-leading} follows immediately. 
\begin{corollary}\label{coro:trivial-leading}
    Let $L$ be an $\omega$-regular language with $|\canoEq| = 1$.
    Then, periodic, syntactic, recurrent and limit FDFAs coincide.
\end{corollary}

We show in Lemma \ref{lem:coarser-than-syntactic} that the limit FDFAs are a coarser representation of $L$ than the syntactic FDFAs.
Moreover, there is a tight connection between the syntactic FDFAs and limit FDFAs.
\begin{lemma}\label{lem:coarser-than-syntactic}
For all $u, x, y \in \finwords$,
\begin{enumerate}
    \item 
    $x \proEq^u_S y$ if, and only if $u\cdot x \canoEq u \cdot y$ and $x \proEq^u_L y$. 
    \item $|\proEq^u_L| \leq |\proEq^u_S| \leq |\canoEq| \cdot |\proEq^u_L|$; $|\proEq^u_L| \leq |\canoEq| \cdot |\proEq^u_P|$.
\end{enumerate}
\end{lemma}
\begin{proof}

\begin{enumerate}
    \item 
    
    \begin{itemize}
        \item Assume that $ux \canoEq uy$ and $x \proEq^u_L y$.
        Since $x \proEq^u_L y$ holds, then for all $v \in \finwords$, $(u x v \canoEq u \implies u \cdot (xv)^{\omega} \in L) \Longleftrightarrow (u y v \canoEq u \implies u \cdot (yv)^{\omega} \in L)$.
        Since $u x \canoEq u y$ holds, then $u \cdot x v \canoEq u \Longleftrightarrow u \cdot y v \canoEq u$ for all $v \in \finwords$.
        Hence, by Definition \ref{def:syntactic-rcs}, if $u xv \not \canoEq u$ (and thus $u yv \not \canoEq u$), it follows that $x \proEq^u_S y$ by definition of $\proEq^u_S$;
        otherwise we have both $u xv \canoEq u$ and $u yv \canoEq u$ hold, and also $u\cdot (xv)^{\omega} \in L\Longleftrightarrow u\cdot (yv)^{\omega} \in L$, following the definition of $\proEq^u_L$.
        It thus follows that $x \proEq_S^u y$.
        \item Assume that $x \proEq^u_S y$.
        First, we have $ux \canoEq u y$ by definition of $\proEq^u_S$.
        Since $u x \canoEq u y$ holds, then $u \cdot x v \canoEq u \Longleftrightarrow u \cdot y v \canoEq u$ for all $v \in \finwords$.
        Assume by contradiction that $x \proEq^u_L y$. Then there must exist some $v \in \finwords$  such that $u \cdot x v \canoEq u \cdot y v \canoEq u$ holds but $u\cdot (xv)^{\omega} \in L \Longleftrightarrow u\cdot (yv)^{\omega} \in L$ does not hold.
        By definition of $\proEq^u_S$, it then follows that $x \not \proEq^S_u y$, violating our assumption.
        Hence, both $ux \canoEq uy$ and $x \proEq^u_L y$ hold.
    \end{itemize}

    \item 
    As an immediate result of the Item (1), we have that  $|\proEq^u_L| \leq |\proEq^u_S| \leq \linebreak |\canoEq| \cdot |\proEq^u_L|$.
    We prove the second claim by showing that, for all $u, x,y \in \finwords$, if $u x \canoEq u y$ and $x \proEq^u_P y$, then $x \proEq^u_S y$ (and thus $x \proEq^u_L y$).
    Fix a word $v \in \finwords$.
    Since $ux \canoEq uy$ holds, it follows that $u x \cdot v \canoEq u \Longleftrightarrow uy\cdot v \canoEq u$.
    Moreover, we have $u\cdot (xv)^{\omega} \in L \Longleftrightarrow u\cdot (yv)^{\omega} \in L$ because $x \proEq^u_P y$ holds.
    By definition of $\proEq_S^u$, it follows that $x \proEq_S^u y$ holds.
    Hence, $x \proEq^u_L y$ holds as well.
    We then conclude that $| \proEq^u_L | \leq |\canoEq| \cdot |\proEq^u_P|$.\qedhere 
    %
\end{enumerate}  
\end{proof}

According to Definition~\ref{def:induced-dfw}, we have $x \canoEq y$ iff $\T[\canoEq](x) = \T[\canoEq](y)$ for all $x, y \in \finwords$.
That is, $\M = (\T[\canoEq], \emptyset)$ is consistent with $\canoEq$, i.e., $x \canoEq y$ iff $\M(x) = \M(y)$ for all $x,y\in \finwords$.
Hence, $u \cdot v \canoEq u$ iff $\M(u) = \M(u \cdot v)$.
In the remaining part of the paper, we may therefore mix the use of $\canoEq$ and $\M$ without distinguishing the two notations. 
    
We are now ready to give our main result of this section.

\begin{theorem}\label{thm:limit-fdfa-rep}
    Let $L$ be an $\omega$-regular language and
    $\F_L {=} (\M[\canoEq], \setnocond{ \N[\proEq_u]}_{\class{u} \in \quotient})$ be the limit FDFA of $L$.
    Then (1) $\F_L$ has a finite number of states,
    (2) $\upword{\F_L} = \upword{L}$, and (3) $\F_L$ is saturated.
\end{theorem}
\begin{proof}
    Since the syntactic FDFA $\F_S$ of $L$ has a finite number of states~\cite{DBLP:journals/tcs/MalerS97} and $\F_L$ is a coarser representation than $\F_S$ (cf.\ Lemma~\ref{lem:coarser-than-syntactic}), $\F_L$ must have finite number of states as well.

    To show $\upword{\F_L} \subseteq \upword{L}$,
    assume that $w \in \upword{\F_L}$.
    By Definition~\ref{def:acc-fdfa}, a UP-word $w$ is accepted by $\F_L$ if there exists a decomposition $(u, v)$ of $w$ such that $\M(u) = \M(u \cdot v)$ (equivalently, $u \cdot v \canoEq u$) and $v \in \finlang{\N^{\rep{u}}_L}$ where $\rep{u} = \M(u)$.
    Here $\rep{u}$ is the representative word for the equivalence class $[u]_{\canoEq}$.
    Similarly, let $\rep{v} = \N^{\rep{u}}_L(v)$.
    By Definition \ref{def:limit-rcs}, we have $\rep{u} \cdot \rep{v} \canoEq \rep{u}\implies \rep{u} \cdot \rep{v}^{\omega} \in L$ holds as $\rep{v}$ is a final state of $\N^{\rep{u}}_L$.
    Since $v \proEq^{\rep{u}}_L \rep{v}$ (i.e., $\N^{\rep{u}}_L(v) = \N^{\rep{u}}_L(\rep{v})$), $\rep{u} \cdot v \canoEq \rep{u}\implies \rep{u} \cdot v^{\omega} \in L$ holds as well.
    It follows that $u \cdot v \canoEq u \implies u \cdot v^{\omega} \in L$ since $u \canoEq \rep{u}$ and $u \cdot v \canoEq \rep{u} \cdot v$ (equivalently, $\M(u \cdot v) = \M(\rep{u} \cdot v)$).
    Together with the assumption that $\M(u \cdot v) = \M(u)$ (i.e, $u \canoEq u \cdot v$), we then have that $u \cdot v^{\omega} \in L$ holds.
    So, $\upword{\F_L} \subseteq \upword{L}$ also holds.
    
    To show that $\upword{L} \subseteq \upword{\F_L}$ holds, let $w \in \upword{L}$.
    For a UP-word $w \in L$, we can find a normalized decomposition $(u, v)$ of $w$ such that $w = u\cdot v^{\omega}$ and $u \cdot v \canoEq u$ (i.e., $\M(u) = \M(u \cdot v)$), since the index of $\canoEq$ is finite (cf.~\cite{DBLP:journals/tcs/AngluinF16} for more details).
    Let $\rep{u} = \M(u)$ and $\rep{v} = \N^{\rep{u}}_L(v)$.
    Our goal is to prove that $\rep{v}$ is a final state of $\N^{\rep{u}}_L$.
    Since $u \canoEq \rep{u}$ and $u\cdot v^{\omega} \in L$, then $\rep{u}\cdot v^{\omega} \in L$ holds.
    Moreover, $\rep{u} \cdot v \canoEq \rep{u}$ holds as well because $\rep{u} = \M(\rep{u}) = \M(u)= \M(\rep{u} \cdot v) = \M(u \cdot v)$.
    (Recall that $\M$ is deterministic.)
    Hence, $\rep{u} \cdot v \canoEq \rep{u} \implies \rep{u}\cdot v^{\omega} \in L$ holds.
    Since $\rep{v} \proEq^{\rep{u}}_L v$, it follows that $\rep{u} \cdot \rep{v} \canoEq \rep{u} \implies \rep{u}\cdot \rep{v}^{\omega} \in L$ also holds.
    Hence, $\rep{v}$ is a final state.
    Therefore, $(u, v)$ is accepted by $\F_L$, i.e., $w \in \upword{\F_L}$.
    It follows that $\upword{L} \subseteq \upword{\F_L}$.

    Now we show that $\F_L$ is saturated.
    Let $w$ be a UP-word.
    Let $(u, v)$ and $(x, y)$ be two normalized decompositions of $w$ with respect to $\M$ (or, equivalently, to $\canoEq$).
    Assume that $(u, v)$ is accepted by $\F_L$.
    From the proof above, it follows that both $u\cdot v \canoEq u $ and $ u\cdot v^{\omega} \in L$ hold.
    So, we know that $u\cdot v^{\omega} = x\cdot y^{\omega} \in L$.
    Let $\rep{x} = \M(x)$ and $\rep{y} = \N^{\rep{x}}_L(y)$.
    Since $(x, y)$ is a normalized decomposition, it follows that $x \cdot y \canoEq x$.
    Again, since $\rep{x} \canoEq x$, $\rep{x} \cdot y \canoEq \rep{x}$ and $\rep{x} \cdot y^{\omega} \in L$ also hold.
    Obviously, $\rep{x}\cdot y \canoEq \rep{x} \implies \rep{x} \cdot y^{\omega} \in L$ holds.
    By the fact that $y \proEq^{\rep{x}}_L \rep{y}$, $\rep{x}\cdot \rep{y} \canoEq \rep{x} \implies \rep{x} \cdot \rep{y}^{\omega} \in L$ holds as well.
    Hence, $\rep{y}$ is a final state of $\N^{\rep{x}}_L$.
    In other words, $(x, y)$ is also accepted by $\F_L$.
    The proof for the case when $(u, v)$ is not accepted by $\F_L$ is similar. 
\end{proof}

\subsection{Size comparison with other canonical FDFAs}
\label{ssec:size-comparison}
As aforementioned, Angluin and Fisman in \cite{DBLP:journals/tcs/AngluinF16} showed that for a variant of the family of languages $L_n$ given by Michel~\cite{michel1988complementation}, its periodic FDFA has $\Omega(n!)$ states, while the syntactic FDFA only has $\bigO(n^2)$ states.
Since limit FDFAs are smaller than syntactic FDFAs, it immediately follows that:
\begin{corollary}
    There exists a family of languages $L_n$ such that its periodic FDFA has $\Omega(n!)$ states, while the limit FDFA only has $\bigO(n^2)$ states.
\end{corollary}

Now we consider the size comparison between limit and recurrent FDFAs.
Consider again the limit and recurrent FDFAs of the language $L = a^{\omega} + ab^{\omega}$ in Figure~\ref{fig:limit-fdfa-example}:
one can see that limit FDFA and recurrent FDFA have the same number of states, even though with different progress DFAs.
In fact, it is easy to see that limit FDFAs and recurrent FDFAs are incomparable regarding the their number of states, even when only the $\omega$-regular languages recognized by weak DBAs are considered.
A \emph{weak} DBA (wDBA) is a DBA in which each SCC contains either all accepting transitions or non-accepting transitions.
\begin{lemma}\label{lem:incomparable-size}
    If $L$ is a wDBA-recognizable language, then its limit FDFA and its recurrent FDFA have incomparable size.
\end{lemma}
\begin{proof}
    We fix $u, x, y \in \finwords$ in the proof.
    Since $L$ is recognized by a wDBA, the TS $\T[\canoEq]$ of the leading DFA $\M$ is isomorphic to the minimal wDBA recognizing $L$~\cite{DBLP:journals/iandc/MalerP95}.
    Therefore, a state $[u]_{\canoEq}$ of $\M$ is either transient, in a rejecting SCC, or in an accepting SCC. We consider these three cases.
    \begin{itemize}
        \item Assume that $[u]_{\canoEq}$ is a transient SCC/state.
        Then for all $v \in \finwords$, $u \cdot x \cdot v \not \canoEq u$ and $u \cdot y \cdot v \not\canoEq u $.

        By the definitions of $\proEq^u_R$
        and $\proEq^u_L$, there are a non-final class $[\emptyword]_{\proEq^u_L}$ and \emph{possibly} a sink final class $[\sigma]_{\proEq^u_L}$ for $\proEq^u_L$ where $\sigma \in \alphabet$, while there is a non-final class $[\emptyword]_{\proEq^u_R}$ for $\proEq^u_R$.
        Hence, $ x \proEq^u_L y$ implies $x \proEq^u_R y$.
        \item Assume that $[u]_{\canoEq}$ is in a rejecting SCC.
        Obviously, for all $v \in \finwords$, we have that $u \cdot x \cdot v \canoEq u \implies u\cdot (x\cdot v)^{\omega} \notin L$ and $u \cdot y \cdot v \canoEq u \implies  u\cdot (y \cdot v)^{\omega} \notin L$.
        Therefore, there is only one equivalence class $[\emptyword]_{\proEq^u_R}$ for $\proEq^u_R$.
        It follows that $ x \proEq^u_L y$ implies $x \proEq^u_R y$.
        \item Assume that $\class{u}$ is in an accepting SCC.
        Clearly, for all $v \in \finwords$, we have that both $u \cdot x \cdot v \canoEq u \implies u\cdot (x\cdot v)^{\omega} \in L$ and $u \cdot y \cdot v \canoEq u \implies  u\cdot (y\cdot v)^{\omega} \in L$ hold.
        That is, we have either $u \cdot x \cdot v \canoEq u \land u\cdot (x\cdot v)^{\omega} \in L$ hold, or $u \cdot x \cdot v \not\canoEq u $.
        If $x \proEq^u_R y$ holds, it immediately follows that $(u \cdot x \cdot v \canoEq u \implies u\cdot (x\cdot v)^{\omega} \in L) \Longleftrightarrow (u \cdot y \cdot v \canoEq u \implies u\cdot (y\cdot v)^{\omega} \in L )$ holds.
        Hence, $x \proEq^u_R y$ implies $x \proEq^u_L y$.
    \end{itemize}
    
    Based on this argument, it is easy to find a language $L$ such that its limit FDFA is more succinct than its recurrent FDFA and vice versa, depending on the size comparison between rejecting SCCs and accepting SCCs. 
    Therefore, the lemma follows.
\end{proof}

\begin{figure}[t]
    \centering
    \scalebox{0.94}{
    \includegraphics{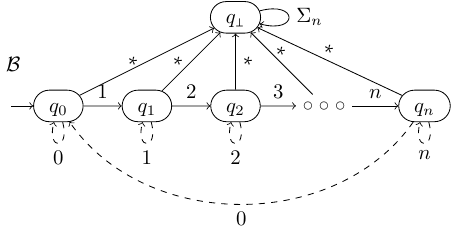}
    }
    \caption{The $\omega$-regular language $L_n$ represented with a DBA $\B$. The dashed arrows are $\accd$-transitions and $*$-transitions represent the missing transitions.}
    \label{fig:dba}
\end{figure}

Lemma~\ref{lem:incomparable-size} reveals that limit FDFAs and recurrent FDFAs are incomparable in size.
Nonetheless, we still provide a family of languages $L_n$ in Lemma~\ref{lem:lower-bound} such that the recurrent FDFA has $\Theta(n^2)$ states, while its limit FDFA only has $\Theta(n)$ states.
One can, of course, obtain the opposite result by complementing $L_n$.
Notably, Lemma~\ref{lem:lower-bound} also gives a matching lower bound for the size comparison between syntactic FDFAs and limit FDFAs, since syntactic FDFAs can be quadratically larger than their limit FDFA counterparts, as stated in Lemma~\ref{lem:coarser-than-syntactic}. 
The language which witnesses this lower bound is given as its DBA $\B$ depicted in Figure~\ref{fig:dba}.
We refer to Appendix~\ref{app:lower-bound} for detailed proof.

\begin{restatable}{lemma}{lemLowBound}\label{lem:lower-bound}
    Let $\alphabet_n = \setnocond{0, 1,\cdots, n}$.
    There exists an $\omega$-regular language $L_n$ over $\alphabet_n$ such that its limit FDFA has $\Theta(n)$ states, while both its syntactic and recurrent FDFAs have $\Theta({n^2})$ states. 
\end{restatable}

Finally, it is time to derive yet another ``Myhill-Nerode" theorem for $\omega$-regular languages, as stated in Theorem~\ref{thm:limit-myhill-nerode}.
This result follows immediately from Lemma~\ref{lem:coarser-than-syntactic} and a similar theorem about syntactic FDFAs~\cite{DBLP:journals/tcs/MalerS97}.
\begin{theorem}\label{thm:limit-myhill-nerode}
    Let $\F_L$ be the limit FDFA of an $\omega$-language $L$.
    Then $L$ is regular if, and only if $\F_L$ has finite number of states.
\end{theorem}
For identifying whether $L$ is DBA-recognizable with FDFAs, a straight forward way as mentioned in the introduction is to go through determinization, which is, however, exponential in the size of the input FDFA. 
We show in Section~\ref{sec:dba-recognizable} that there is a polynomial-time algorithm using our limit FDFAs.

\section{Limit FDFAs for identifying DBA-recognizable languages}
\label{sec:dba-recognizable}

Given an $\omega$-regular language $L$, we show in this section how to use the limit FDFA of $L$ to check whether $L$ is DBA-recognizable in polynomial time. 
To this end, we will first introduce how the limit FDFA of $L$ looks like in Section~\ref{ssec:dba-limit-fdfa} and then introduce the deciding algorithm in Section~\ref{ssec:deciding-algo}.

\subsection{Limit FDFA for DBA-recognizable languages}
\label{ssec:dba-limit-fdfa}

Bohn and L\"oding~\cite{DBLP:conf/icalp/BohnL22} construct a type of family of DFAs $\F_{BL}$ from a set $S^+$ of positive samples and a set $S^-$ of negative samples, where the progress DFA accepts exactly the language $V_u = \setnocond{x \in \poswords: \forall v \in \finwords.\ \text{if } u\cdot xv \canoEq u, \text{ then } u \cdot (xv)^{\omega} \in L}$\footnote{Defining directly a progress RC $\proEq^u$ that recognizes $V_u$ is hard since $V_u$ is quantified over all $v$-extensions.}.
When the samples $S^+$ and $S^-$ uniquely characterize a DBA-recognizable language $L$, $\F_{BL}$ recognizes exactly $L$.

The progress DFA $\N^u_L$ of our limit FDFA $\F_L$ of $L$ usually accepts \emph{more} words than $V_u$.
Nonetheless, we can still find one final equivalence class that is exactly the set $V_u$, as stated in Lemma~\ref{lem:dba-lang-co-safety}.
\begin{restatable}{lemma}{dbaCoSafe}\label{lem:dba-lang-co-safety}
    Let $L$ be a DBA-recognizable language and $\F_L {=} (\M, \setnocond{\N^u_L}_{\class{u}\in \finwords/_{\canoEq}})$ be the limit FDFA of $L$.
    Then, for each progress DFA $\N^u_L$ with $\finlang{\N^u_L} \neq \emptyset$, there must exist a final state $\rep{x} \in F_{u}$ such that $[\rep{x}]_{\proEq^u_L} = \setnocond{ x \in \poswords: \forall v \in \finwords.\ u \cdot (x \cdot v) \canoEq u \implies u\cdot (x\cdot v)^{\omega} \in L}$.
\end{restatable}

\begin{proof}
    In~\cite{DBLP:conf/icalp/BohnL22}, it is shown that for each equivalence class $[u]_{\canoEq}$ of $\canoEq$, there exists a regular language $V_u = \setnocond{x \in \poswords: \forall v \in \finwords.\ \text{if } u\cdot xv \canoEq u, \text{ then } u \cdot (xv)^{\omega} \in L}$.
    We have also provided the proof of the existence of $V_u$ in Appendix~\ref{app:dba-cosafety}, adapted to our notations.
    The intuition of $V_u$ is the following.
    Let $\B = (\alphabet, \states, \iota, \trans, \accd)$ be a DBA accepting $L$.
    Then, $[u]_{\canoEq}$ corresponds to a set of states $S = \setnocond{q \in \states: q = \trans(\iota, u'), u' \in [u]_{\canoEq}}$ in $\B$.
    For each $q \in S$, we can easily create a regular language $V_q$ such that $x \in V_q$ iff over the word $x$, $\B^q$ (the DBA derived from $\B$ by setting $q$ its initial state) visits an accepting transition, $\B^q$ goes to an SCC that cannot go back to $q$, or $\B^q$ goes to a state that cannot go back to $q$ unless visiting an accepting transition.
    Then, $V_u = \cap_{q \in S} V_q$.
    
    Now we show that $V_u$ is an equivalence class of $\proEq^u_L$ as follows.
    On one hand, for every two different words $x_1, x_2 \in V_u$, we have that $x_1 \proEq^u_L x_2$, which is obvious by the definition of $V_u$.
    On the other hand, it is easy to see that $x' \not\proEq^u_L x$ for all $x' \notin V_u$ and $x \in V_u$ because there exists some $v \in \finwords$ such that $u \cdot x' \cdot v \canoEq u$ but $u\cdot (x' \cdot v)^{\omega} \notin L$.
    Hence, $V_u$ is indeed an equivalence class of $\proEq^u_L$.
    Obviously, $V_u \subseteq \finlang{\N^u_L}$, as we can let $v = \emptyword$, so for every word $x \in V_u$, we have that $u \cdot x \canoEq u \implies u \cdot x^{\omega} \in L$.
    Let $\rep{x} = \N^u_L(x)$ for a word $x \in V_u$.
    It follows that $\rep{x}$ is a final state of $\N^u_L$ and we have $[\rep{x}]_{\proEq^u_L} = V_u$.
    This completes the proof.
\end{proof}

By Lemma~\ref{lem:dba-lang-co-safety}, we can define a variant of limit FDFAs for only DBAs with less number of final states.
This helps to reduce the complexity when translating FDFAs to NBAs~\cite{DBLP:conf/mfps/CalbrixNP93,AngluinBF18,DBLP:journals/iandc/LiCZL21}.
Let $n$ be the number of states in the leading DFA $\M$ and $k$ be the number of states in the largest progress DFA.
Then the resultant NBA from an FDFA has $\bigO(n^2k^3)$ states~\cite{DBLP:conf/mfps/CalbrixNP93,AngluinBF18,DBLP:journals/iandc/LiCZL21}.
However, if the input FDFA is $\F_B$ as in Definition~\ref{def:limit-rcs-dba}, the complexity of the translation will be $\bigO(n^2k^2)$, as there is at most one final state, rather than $k$ final states, in each progress DFA.

\begin{definition}[Limit FDFAs for DBAs]\label{def:limit-rcs-dba}
The limit FDFA $\F_B = (\M, \setnocond{\N^{u}_B})$ of $L$ is defined as follows.

The transition systems of $\M$ and $\N^u_B$ for each $[u]_{\canoEq} \in \finwords/_{\canoEq}$ are exactly the same as in Definition~\ref{def:limit-rcs}.

The set of final states $F_{u}$ contains the equivalence classes $[x]_{\proEq^u_L}$ such that, for all $v \in \finwords$, $u \cdot xv \canoEq u \Longrightarrow u\cdot (xv)^{\omega} \in L$ holds.
\end{definition}

The change to the definition of final states would not affect the language that the limit FDFAs recognize, but only their saturation properties.
We say an FDFA $\F$ is \emph{almost saturated} if, for all $u, v \in \finwords$, we have that if $(u, v)$ is accepted by $\F$, then $(u, v^k)$ is accepted by $\F$ for all $k \geq 1$. 
According to~\cite{DBLP:journals/iandc/LiCZL21}, if $\F$ is almost saturated, then the translation algorithm from FDFAs to NBAs in~\cite{DBLP:conf/mfps/CalbrixNP93,AngluinBF18,DBLP:journals/iandc/LiCZL21} still applies (cf. Appendix~\ref{app:fdfa-to-nba} about details of the NBA construction).

\begin{theorem}\label{thm:dfa-limit-lang}
    Let $L$ be a DBA-recognizable language and $\F_B$ be the limit FDFA induced by Definition~\ref{def:limit-rcs-dba}.
    Then (1) $\upword{\F_B} = \upword{L}$ and (2) $\F_B$ is almost saturated but not necessarily saturated.
\end{theorem}
\begin{proof}
    The proof for $\upword{\F_B} \subseteq \upword{L}$ is trivial, as the final states defined in Definition~\ref{def:limit-rcs-dba} must also be final in Definition~\ref{def:limit-rcs}.
    The other direction can be proved based on Lemma~\ref{lem:dba-lang-co-safety}.
    Let $w \in \upword{L}$ and $\B = (\states, \alphabet, \iota, \trans, \accd)$ be a DBA accepting $L$.
    Let $\rho$ be the run of $\B$ over $w$.
    We can find a decomposition $(u, v)$ of $w$ such that there exists a state $q$ with $q = \trans(\iota, u) = \trans(\iota, u \cdot v)$ and $(q, v[0]) \in \accd$.
    As in the proof of Lemma~\ref{lem:dba-lang-co-safety}, we are able to construct the regular language $V_u = \setnocond{x \in \poswords: \forall y \in \finwords, u\cdot x \cdot y \canoEq u\implies u \cdot (x\cdot y)^{\omega} \in L}$.
    We let $S = \setnocond{p \in \states: \lang{\B^q} = \lang{\B^p}}$.
    For every state $p \in S$, we have that $v^{\omega} \in \lang{\B^p}$.
    For each $p \in S$, we select an integer $k_p > 0$ such that the finite run $p \xrightarrow{v^{k_p}} \trans(p, v^{k_p})$ visits some accepting transition.
    Then we let $k = \max_{p \in S} k_p$.
    By definition of $V_u$, it follows that $v^{k} \in V_u$.
    That is, $V_u$ is not empty.
    According to Lemma~\ref{lem:dba-lang-co-safety}, we have a final equivalence class $[x]_{\proEq^u_L} = V_u$ with $v^k \in [x]_{\proEq^u_L}$.
    Moreover, $u \cdot v^k \canoEq u$ since $q = \trans(\iota, u) = \trans(q, v)$.
    Hence, $(u, v^k)$ is accepted by $\F_B$, i.e., $w \in \upword{\F_B}$.
    It follows that $\upword{\F_B} = \upword{L}$.
    
    Now we prove that $\F_B = (\M, \setnocond{\N^u_B})$ is \emph{not} necessarily saturated.
    Let $L = (\finwords \cdot aa)^{\omega}$.
    Obviously, $L$ is DBA recognizable,
    and $\canoEq$ has only one equivalence class, $[\emptyword]_{\canoEq}$.
    Let $ w = a^{\omega} \in \upword{L}$.
    Let $(u= \emptyword, v = a)$ be a normalized decomposition of $w$ with respect to $\canoEq$ (thus, $\M$).
    We can see that there exists a finite word $x$ (e.g., $x=b$ is such a word) such that $ \emptyword \cdot a \cdot x \canoEq \emptyword$ and $\emptyword \cdot (a \cdot x)^{\omega} \notin L$.
    Thus, $(\emptyword, a)$ will not be accepted by $\F_B$.
    Hence $\F_B$ is not saturated.
    Nonetheless, it is easy to verify that $\F_B$ is almost saturated.
    Assume that $(u, v)$ is accepted by $\F_B$.
    Let $\rep{u} = \M(u)$ and $\rep{v} = \N^{\rep{u}}_B(v)$.
    Since $\rep{v}$ is the final state, then, according to Definition~\ref{def:limit-rcs-dba},  we have for all $e \in \finwords$ that $\rep{u} \cdot \rep{v} e \canoEq \rep{u} \implies \rep{u} \cdot (\rep{v} e)^{\omega} \in L$.
    Since $v \proEq^u_L \rep{v}$, $\rep{u} \cdot v e \canoEq \rep{u} \implies \rep{u} \cdot (v e)^{\omega} \in L$ also holds for all $e \in \finwords$.
    Let $e = v^k\cdot e'$ where $e' \in \finwords, k \geq 0$.
    It follows that $\rep{u} \cdot v^k e' \canoEq \rep{u} \implies \rep{u} \cdot (v^k e')^{\omega} \in L$ holds for $k \geq 1$ as well. 
    Therefore, for all $e' \in \finwords, k \geq 1$, $(\rep{u} \cdot \rep{v} e' \canoEq \rep{u} \implies \rep{u} \cdot (\rep{v} e')^{\omega} \in L ) \Longleftrightarrow (\rep{u} \cdot v^k e' \canoEq \rep{u} \implies \rep{u} \cdot (v^k e')^{\omega} \in L )$ holds.
    In other words, $\rep{v} \proEq^{\rep{u}}_L v^k$ for all $k \geq 1$.
    Together with that $u v^k\canoEq u$, $(u, v^k)$ is accepted by $\F_B$ for all $k \geq 1$.
    Hence, $\F_B$ is almost saturated.
\end{proof}





\subsection{Deciding DBA-recognizable languages}
\label{ssec:deciding-algo}

We show next how to identify whether a language $L$ is DBA-recognizable with our limit FDFA $\F_L$.
Our decision procedure relies on the translation of FDFAs to NBAs/DBAs.
In the following, we let $n$ be the number of states in the leading DFA $\M$ and $k$ be the number of states in the largest progress DFA.
We first give some previous results below.

\begin{lemma}[\hspace*{-1mm}{\cite[Lemma 6]{DBLP:journals/iandc/LiCZL21}}]\label{lem:translation}
    Let $\F$ be an (almost) saturated FDFA of $L$. Then one can construct an NBA $\A$ with $\bigO(n^2k^3)$ states such that $\lang{\A} = L$.
\end{lemma}

Now we consider the translation from FDFA to DBAs.
By Lemma~\ref{lem:dba-lang-co-safety}, there is a final equivalence class $[x]_{\proEq^u_L}$ that is a \emph{co-safety} language in the limit FDFA of $L$.
Co-safety regular languages are regular languages $R \subseteq \finwords$ such that $R \cdot \finwords = R$.
It is easy to verify that if $x' \in [x]_{\proEq^u_L}$, then $x'v\in [x]_{\proEq^u_L}$ for all $v\in \finwords$, based on the definition of $\proEq^u_L$.
So, $[x]_{\proEq^u_L}$ is a co-safety language.
The DFAs accepting co-safety languages usually have a sink final state $f$ (such that $f$ transitions to itself over all letters in $\alphabet$). 
We therefore have the following.

\begin{corollary}\label{coro:necessary-condition}
    If $L$ is DBA-recognizable then every progress DFA $\N^u_L$ of the limit FDFA $\F_L$ of $L$ either has a sink final state, or no final state at all.
\end{corollary}

Our limit FDFA $\F_B$ of $L$, as constructed in Definition~\ref{def:limit-rcs-dba}, accepts the same co-safety languages in the progress DFAs as the FDFA obtained in \cite{DBLP:conf/icalp/BohnL22}, although they may have different transition systems.
Nonetheless, we show that their DBA construction still works on $\F_B$.
To make the construction more general, we assume an FDFA $\F = (\M, \setnocond{\N^q}_{q \in \states})$ where $\M = (\states, \alphabet, \iota, \trans)$ and, for each $q \in \states$, we have $\N^q = (\states_q,  \alphabet,\iota_{q}, \trans_q, F_q)$.
\begin{definition}[\hspace*{-1.3mm}\cite{DBLP:conf/icalp/BohnL22}]\label{def:dba-construction}
    Let $\F =  (\M, \setnocond{ \N^q}_{q \in \states})$ be an FDFA.
    Let $\T[\F]$ be the TS constructed from $\F$ defined as the tuple  $\T[\F] = (\states_{\T}, \alphabet, \iota_{\T}, \trans_{\T})$ and $\accd \subseteq \setnocond{(q, \sigma): q \in \states_{\T}, \sigma\in \alphabet}$ be a set of transitions where
\begin{itemize}
    \item $Q_{\T} := \states \times  \bigcup_{q\in\states} \states_q$;
    \item $\iota_{\T} := (\iota, \iota_{\iota})$;
    \item For a state $(m, q) \in Q_{\T}$ and $\sigma \in \alphabet$, let $q' = \trans_{\widetilde{m}}(q, \sigma)$ where $\N^{\widetilde{m}}$ is the progress DFA that $q$ belongs to and let $m' = \trans(m, \sigma)$.
    Then
\begin{equation*}
\trans((m, q), \sigma) =
    \begin{cases}
        (m', q') & \text{if }  q' \notin F_{\widetilde{m}}\\
        (m', \iota_{m'}) & \text{if } q' \in F_{\widetilde{m}}
    \end{cases}
\end{equation*}
    \item $((m, q), \sigma) \in \accd$ if $q' \in F_{\widetilde{m}}$ 
\end{itemize}
\end{definition}

\begin{lemma}\label{lem:dba-lower-approx}
    If $\F$ is an FDFA with only sink final states.
    Let $\B[\F] = (\T[\F], \accd)$ as given in Definition~\ref{def:dba-construction}.
    Then, $\upword{\lang{\B[\F]}} \subseteq \upword{\F}$.
\end{lemma}
\begin{proof}
    Let $w \in \upword{\lang{\B[\F]}}$ and $\rho$ be its corresponding accepting run.
    Since $w$ is a UP-word and $\B[\F]$ is a DBA of finite states, then we must be able to find a decomposition $(u, v)$ of $w$ such that $(m, \iota_m) = \B[\F](u) = \B[\F](u \cdot v)$, where $\rho$ will visit a $\accd$-transition whose destination is $(m, \iota_m)$ for infinitely many times.
    It is easy to see that $\M(u \cdot v) =\M( u)$ since $\B[\F](u) = \B[\F](u \cdot v)$.
    Moreover, we can show there must be a prefix of $v$, say $v'$, such that $v' \in \finlang{\N^m}$.
    Since $\finlang{\N^m}$ is co-safety, we have that $v \in \finlang{\N^m}$.
    Thus, $(u, v)$ is accepted by $\F$.
    By Definition~\ref{def:acc-fdfa}, $w \in \upword{\F}$.
    Therefore, $\upword{\lang{\B[\F]}} \subseteq \upword{\F}$.    
\end{proof}

\begin{figure}
    \centering
    \scalebox{0.94}{
    \includegraphics{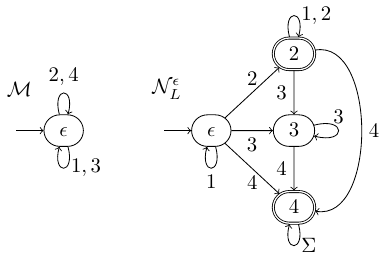}
    }
    \caption{An example limit FDFA $\F = (\M, \setnocond{\N^{\emptyword}_L})$}
    \label{fig:example-limit-fdfa}
\end{figure}

By Corollary~\ref{coro:necessary-condition}, $\F_B$ has only sink final states;
so, we have that $\upword{\lang{\B[\F_B]}} \subseteq \upword{\F_B}$. 
However, Corollary~\ref{coro:necessary-condition} is only a necessary condition for $L$ being DBA-recognizable, as explained below.
Let $L $ be an $\omega$-regular language over $\alphabet = \setnocond{1, 2, 3, 4}$ such that a word $w \in L$ iff the maximal number that occurs infinitely often in $w$ is even.
Clearly, $L$ has one equivalence class $[\emptyword]_{\canoEq}$ for $\canoEq$.
The limit FDFA $\F = (\M, \setnocond{\N^{\emptyword}_L})$ of $L$ is depicted in Figure~\ref{fig:example-limit-fdfa}.
We can observe that the equivalence class $[4]_{\proEq^{\emptyword}_L}$ corresponds to a co-safety language.
Hence, the progress DFA $\N^{\emptyword}_L$ has a sink final state.
However, $L$ is not DBA-recognizable.
If we ignore the final equivalence class $[2]_{\proEq^{\emptyword}_L}$ and obtain the variant limit FDFA $\F_B$ as given in Definition~\ref{def:limit-rcs-dba}, then we have $\upword{\F_B} \neq \upword{L}$ since the $\omega$-word $2^{\omega}$ is missing.
But then, by Theorem~\ref{thm:dfa-limit-lang}, this change would not lose words in $L$ if $L$ is DBA-recognisable, leading to contradiction.
Therefore, $L$ is shown to be not DBA-recognizable.
So the key of the decision algorithm here is to check whether ignoring other final states will retain the language.
With Lemma~\ref{lem:dba-completeness}, we guarantee that $\B[\F_B]$ accepts exactly $L$ if $L$ is DBA-recognizable.

\begin{lemma}\label{lem:dba-completeness}
    Let $L$ be a DBA-recognizable language.
    Let $\F_B$ be the limit FDFA $L$, as constructed in Definition~\ref{def:limit-rcs-dba}.
    Let $\B[\F_B] = (\T[\F_B], \accd)$, where $\T[\F_B]$ and $\accd$ are the TS and set of transitions, respectively, defined in Definition~\ref{def:dba-construction} from $\F_B$.
    Then $\upword{\F_B} = \upword{L} \subseteq 
 \upword{\lang{\B[\F_B]}} $.
\end{lemma}
\begin{proof}
    We first assume for contradiction that some $w \in L$ is rejected by $\B[\F_B]$.
    For this, we consider the run $\rho = (q_0,w[0],q_1)(q_1,w[1],q_2)\ldots$ of $\B[\F_B]$ on $w$. Let $i \in \omega$ be such that $(q_{i-1},w[i-1],q_i)$ is the last accepting transition in $\rho$, and $i=0$ if there is no accepting transition at all in $\rho$.
    We also set $u=w[0\cdots i-1]$ and $w' = w[i\cdots]$.
    By Definition~\ref{def:dba-construction}, this ensures that $\B[\F_B]$ is in state $([u]_{\canoEq},\iota_{[u]_{\canoEq}})$ after reading~$u$ and will not see accepting transitions (or leave $\mathcal N^{[u]_{\canoEq}}_B$) while reading the tail~$w'$.

    Let $\mathcal D = (Q',\Sigma,\iota',\delta',\Gamma')$ be a DBA that recognizes $L$ and has only reachable states.
    As $\mathcal D$ recognizes $L$, it has the same right congruences as $L$; by slight abuse of notation, we refer to the states in $Q'$ that are language equivalent to the state reachable after reading $u$ by $[u]_{\canoEq}$ and note that $\mathcal D$ is in some state of $[u]_{\canoEq}$ after (and only after) reading a word $u' \canoEq u$.

    As $u\cdot w'$, and therefore $u' \cdot w'$ for all $u' \canoEq u$, are in $L$, they are accepted by $\mathcal D$, which in particular means that, for all $q \in [u]_{\canoEq}$, there is an $i_{q}$ such that there is an accepting transition in the first $i_q$ steps of the run of 
    $\D^{q}$ (the DBA obtained from $\D$ by setting the initial state to $q$) on $w'$. Let $i_+$ be maximal among them and $v=w[i \cdots i+i_+]$.
    Then, for $u' \canoEq u$ and any word $u' v v'$, we either have $u' v v' \not\canoEq u$, or $u' v v' \canoEq u$ \emph{and} $u' \cdot (v v')^\omega \in L$. (The latter is because $v$ is constructed such that a run of $\mathcal D$ on this word will see an accepting transition while reading each $v$, and thus infinitely many times.)
    Thus, $\mathcal N^{[u]_{\canoEq}}_B$ will accept any word that starts with $v$, and therefore be in a final sink after having read $v$.

    But then $\B[\F_B]$ will see another accepting transition after reading $v$ (at the latest after having read $uv$), which closes the contradiction and completes the proof.
\end{proof}

So, our decision algorithm works as follows.
Assume that we are given the limit FDFA $\F_L = (\M, \setnocond{\N^q_L})$ of $L$.
\begin{enumerate}
    \item We first check whether there is a progress DFA $\N^q_L$ such that there are final states but without the sink final state.
    If it is the case, we terminate and return ``NO".
    \item Otherwise, we obtain the FDFA $\F_B$ by keeping the sink final state as the sole final state in each progress DFA (cf. Definition~\ref{def:limit-rcs-dba}).
    Let $\A = \fn{NBA}(\F_L)$ be the NBA constructed from $\F_L$ (cf. Lemma~\ref{lem:translation}) and $\B= \fn{DBA}(\F_B)$ be the DBA constructed from $\F_B$ (cf. Definition~\ref{def:dba-construction}).
    Obviously, we have that $\upword{\lang{\A}} = \upword{L}$ and $\upword{\lang{\B}} \subseteq \upword{\F_B} = \upword{L}$.
    \item Then we check whether $\lang{\A} \subseteq \lang{\B}$ holds.
    If so, we return ``YES", and otherwise ``NO".
\end{enumerate}

Now we are ready to give the main result of this section.
\begin{theorem}
    Deciding whether $L$ is DBA-recognizable can be done in time polynomial in the size of the limit FDFA of $L$.
\end{theorem}
\begin{proof}
We first prove our decision algorithm is correct.
If the algorithm returns ``YES", clearly, we have $\lang{\A} \subseteq \lang{\B}$.
It immediately follows that $\upword{L} = \upword{\lang{\A}} \subseteq \upword{\lang{\B}} \subseteq  \upword{ \F_B}\subseteq \upword{\F_L} = \upword{L}$ according to Lemmas~\ref{lem:translation} and~\ref{lem:dba-lower-approx}.
Hence, $\upword{\lang{\B}} = \upword{L}$, which implies that $L$ is DBA-recognizable.
For the case that the algorithm returns ``NO", we analyze two cases:
\begin{enumerate}
    \item $\F$ has final states but without sink accepting states for some progress DFA. By Corollary~\ref{coro:necessary-condition}, $L$ is not DBA-recognizable.
    \item $\lang{\A} \not\subseteq \lang{\B}$. It means that $\upword{L} \not \subseteq \upword{\lang{\B}}$ (by Lemma~\ref{lem:translation}).
    It follows that $L$ is not DBA-recognizable by Lemma~\ref{lem:dba-completeness}.
\end{enumerate}
The algorithm is therefore sound; its
completeness follows from Lemmas~\ref{lem:dba-lower-approx} and~\ref{lem:dba-completeness}.

The translations above are all in polynomial time.
Moreover, checking the language inclusion between an NBA and a DBA can also be done in polynomial time~\cite{DBLP:journals/jcss/Kurshan87}.
Hence, the deciding algorithm is also in polynomial time in the size of the limit FDFA of $L$.
\end{proof}

Recall that, our limit FDFAs are dual to recurrent FDFAs.
One can observe that, for DBA-recognizable languages, recurrent FDFAs do not necessarily have sink final states in progress DFAs.
For instance, the $\omega$-regular language $L = a^{\omega} + ab^{\omega}$ is DBA-recognizable, but its recurrent FDFA, depicted in Fig.~\ref{fig:limit-fdfa-example}, does not have sink final states.
Hence, our deciding algorithm does not work with recurrent FDFAs.

\section{Underspecifying progress right congruences}
\label{sec:progress-RP}
Recall that recurrent and limit progress DFAs $\N^u$ either treat don't care words in $\overline{C_u} = \setnocond{v\in \poswords: u v \not \canoEq u}$ as rejecting or accepting, whereas it really does not matter whether or not they are accepted.
So why not keep this question open?
We do just this in this section; however, we find that treating the progress with maximal flexibility comes at a cost: the resulting right progress relation $\proEq^u_N$ is \emph{no} longer an equivalence relation, but only a reflexive and symmetric relation over $\finwords \times \finwords$ such that $x \proEq^u_N y$ implies $xv \proEq^u_N yv$ for all $u, x, y, v\in \finwords$.

For this, we first introduce \emph{Right Pro-Congruences} (RP) as relations on words that satisfy all requirements of an RC except for transitivity.

\begin{definition}[Progress RP]\label{def:nondet-proeq}
    Let $[u]_{\canoEq}$ be an equivalence class of $\canoEq$.
    For $x, y \in \finwords$, we define the progress RP $\proEq^u_N$ as follows:
    \[ x \proEq^u_N y \text{ iff } \forall v\in\finwords.\ (u xv \canoEq u \land u yv \canoEq u) \implies (u\cdot (xv)^{\omega} \in L \Longleftrightarrow u\cdot (yv)^{\omega} \in L).\]
\end{definition}
Obviously, $\proEq^u_N$ is a RP, i.e., for $x, y, v'\in\infwords$, if $x \proEq^u_N y$, then $xv' \proEq^u_N yv'$.
That is, assume that $x \proEq^u_N y$ and we want to prove that, for all $e \in \finwords$, $(u \cdot x v' e \canoEq u \land u \cdot y v' e \canoEq u) \implies (u \cdot (x v' e)^{\omega} \in L \Longleftrightarrow u \cdot (yv'e)^{\omega} \in L)$.
This follows immediately by setting $v = v' e$ in Definition \ref{def:nondet-proeq} for all $e \in \finwords$ since $x \proEq^u_N y$.
As $\proEq^u_N$ is not necessarily an equivalence relation%
\footnote{%
In the language $L = a^{\omega} + ab^{\omega}$ from the example of Figure \ref{fig:limit-fdfa-example}, for example, we have $a \proEq^{ab}_N \emptyword$ and $a \proEq^{ab}_N b$, but $b \not\proEq^{ab}_N \emptyword$.}, so that we cannot argue directly with the size of its index.
However, we can start with showing that $\proEq^u_N$ is coarser than $\proEq^u_P, \proEq^u_S, \proEq^u_R$, and $\proEq^u_L$.

\begin{lemma}
    For $u, x, y \in \finwords$, we have that if $x \proEq^u_K y$, then $x \proEq^u_N y$, where $K \in \setnocond{P, S, R, L}$.
\end{lemma}
\begin{proof}
    First, if $x \proEq^u_P y$, $x \proEq^u_N y$ holds trivially.

For syntactic, recurrent, and limit RCs, we first argue for fixed $v \in \finwords$ that
\begin{itemize}
    \item $ux \canoEq uy \Longrightarrow uxv \canoEq uyv$, and therefore

    $ux \canoEq uy \land  \big( u\cdot x \cdot v \canoEq u \Longrightarrow (u\cdot (x \cdot v)^{\omega} \in L \Longleftrightarrow u\cdot (y\cdot v)^{\omega} \in L)\big)$
    
    $\models  (u xv \canoEq u \land u yv \canoEq u) \implies (u\cdot (xv)^{\omega} \in L \Longleftrightarrow u\cdot (yv)^{\omega} \in L)$,
    
    \item 
    $(u \cdot x \cdot v \canoEq u \land u\cdot (x v)^{\omega} \in L) \Longleftrightarrow (u\cdot y v \canoEq u \land u\cdot (y \cdot v)^{\omega} \in L)$
    
    $\models  (u xv \canoEq u \land u yv \canoEq u) \implies (u\cdot (xv)^{\omega} \in L \Longleftrightarrow u\cdot (yv)^{\omega} \in L)$, and
    
    \item  $(u \cdot x \cdot v \canoEq u \Longrightarrow u\cdot (x \cdot v)^{\omega} \in L) \Longleftrightarrow (u \cdot y\cdot v \canoEq u \Longrightarrow u\cdot (y \cdot v)^{\omega} \in L)$
    
    $\models  (u xv \canoEq u \land u yv \canoEq u) \implies (u\cdot (xv)^{\omega} \in L \Longleftrightarrow u\cdot (yv)^{\omega} \in L)$,
\end{itemize}
which is simple Boolean reasoning.
As this holds for all $v \in \finwords$ individually, it also holds for the intersection over all $v \in \finwords$, so that the claim follows.
\end{proof}
Now, it is easy to see that we can use any RC $\proEq$ that refines $\proEq^u_N$ and use it to define a progress DFA.
It therefore makes sense to define the set of RCs that refine $\proEq^u_N$ as $\mathsf{RC}(\proEq^u_N) = \{\proEq \ \mid \ \proEq \subset \proEq^u_N$ is a RC$\}$, and the
best index $|\proEq^u_N|$ of our progress RP as $|\proEq^u_N| = \min\{|\proEq| \ \mid \ \proEq \in \mathsf{RC}(\proEq^u_N)\}$. 
With this definition,
Corollary~\ref{coro:smallest-proEq} follows immediately.
\begin{corollary}\label{coro:smallest-proEq}
    For $u\in \finwords$, we have that $|\proEq^u_N| \leq |\proEq^u_K|$ for all $K \in \setnocond{P, S, R, L}$.
\end{corollary}

We note that the restriction of $\proEq^u_N$ to $C_u \times C_u$ is still an equivalence relation, where  $C_u = \setnocond{v\in\finwords: u v \canoEq u}$ are the words the FDFA acceptance conditions really care about.
This makes it easy to define a DFA over each $\proEq \in \mathsf{RC}(\proEq_N^u)$ with finite index: $C_u/_{\proEq_N^u}$ is good if it contains a word $v$ s.t.\ $u\cdot v^\omega \in L$, and a quotient of $\Sigma^*/_\proEq$ is accepting if it intersects with a good quotient (note that it intersects with at most one quotient of $C_u$).
With this preparation, we now show the following.

\begin{restatable}{theorem}{thmLastFdfaRep}\label{thm:last-fdfa-rep}
    Let $L$ be an $\omega$-regular language and
    $\F_L {=} (\M[\canoEq], \setnocond{ \N[\proEq_u]}_{\class{u} \in \quotient})$ be the limit FDFA of $L$ s.t.\ $\proEq_u \in \mathsf{RC}(\proEq_N^u)$ with finite index for all $u$.
    Then (1) $\F_L$ has a finite number of states,
    (2) $\upword{\F_L} = \upword{L}$, and (3) $\F_L$ is saturated.
\end{restatable}

The proof is similar to the proof of Theorem \ref{thm:limit-fdfa-rep} and moved to Appendix \ref{app:proofofT6}.

\section{Discussion and future work}

Our limit FDFAs fit nicely into the learning framework for FDFAs~\cite{DBLP:journals/tcs/AngluinF16} and are already available for use in the learning library \textsf{ROLL}\footnote{\url{https://github.com/iscas-tis/roll-library}}~\cite{DBLP:conf/tacas/LiSTCX19}.
Since one can treat an FDFA learner as comprised of a family of DFA learners in which one DFA of the FDFA is learned by a separate DFA learner, we only need to adapt the learning procedure for progress DFAs based on our limit progress RCs, without extra development of the framework; see Appendix~\ref{app:learning} for details.
We leave the empirical evaluation of our limit FDFAs in learning $\omega$-regular languages as future work.  

We believe that limit FDFAs are complementing the existing set of canonical FDFAs, in terms of recognizing and learning $\omega$-regular languages.
Being able to easily identify DBA-recognizable languages, limit FDFAs might be used in a learning framework for DBAs using membership and equivalence queries.
We leave this to future work.
Finally, we have looked at retaining maximal flexibility in the construction of FDFA by moving from progress RCs to progress RPs.
While this reduces size, it is no longer clear how to construct them efficiently, which we leave as a future challenge.

\subsubsection*{Acknowledgements}
We thank the anonymous reviewers for their valuable feedback.
This work has been supported by the EPSRC through grants EP/X021513/1 and EP/X017796/1.

\newpage
\bibliographystyle{splncs04}
\bibliography{main}

\clearpage
\appendix

\section{Proof of Lemma \ref{lem:lower-bound}}
\label{app:lower-bound}
\lemLowBound*
\begin{proof}
    The language $L_n$ is given as its DBA $\B = (\states, \alphabet_n, q_0, \trans,\accd)$ depicted in Figure~\ref{fig:dba}, where $\Sigma_n = \{0, 1, \ldots, n\}$.
    First, we show that the index of $\canoEq_{L_n}$ is $n+2$.
    Here we add the subscript $L_n$ to $\canoEq_{L_n}$ to distinguish it from $\canoEq$ for the language $L$.
    In fact, the leading DFA induced by $\canoEq_{L_n}$ is the exactly the TS of $\B$.
    Here, we only show that the limit FDFA and the recurrent FDFA of $L_n$, respectively, have $\Theta(n)$ states and $\Theta(n^2)$ states.
    For every two words $u_1, u_2 \in \finwords$, if $u_1 \not\canoEq_{L_n} u_2$, then there exists a word $w \in \infwords$ such that $u_1 \cdot w \in L_n \Longleftrightarrow u_2 \cdot w \in L_n$ does not hold.
    That is, $u^{-1}_1\cdot L_n \neq u^{-1}_2 \cdot L_n$ where $u^{-1}\cdot L_n = \setnocond{w \in \infwords: u \cdot w \in L_n}$ for a word $u \in \finwords$.
    Let $L_{q} = \lang{\B^q}$.
    For every pair of different states $q_i, q_j \in \states$ with $i \neq j$, obviously $L_{q_i} \neq L_{q_j}$ since $L_{q_i}$ contains an infinite word $i^{\omega}$, while $L_{q_j}$ does not contain such a word. 
    So, if $\B(u_1) \neq \B(u_2)$, then $u^{-1}_1\cdot L_n \neq u^{-1}_2 \cdot L_n$.
    Hence, $|\canoEq_{L_n}| \geq n + 2$.
    It is trivial to see that $|\canoEq_{L_n}| \leq n + 2$ since the index of $\canoEq_{L_n}$ is always not greater than the number of states in a deterministic $\omega$-automaton accepting $L_n$.
    Therefore, $|\canoEq_{L_n}| = n + 2$.

    Now we fix a word $u$ and consider the index of $\proEq^u_L$.
    Let $x \in \finwords$.
    Obviously, if $q_{\bot} = \B(u)$, then for all $v \in \finwords$, we have $u \cdot x \cdot v \canoEq_{L_n} u$ but $u \cdot (x\cdot v)^{\omega} \notin L_n$.
    Hence, $|\proEq^{u}_L | = 1$.
    Now let $q_i = \B(u)$ with $0 \leq i \leq n$.
    For all $v \in \finwords$, if $u \cdot x \cdot v \canoEq_{L_n} u$ holds, it must be the case that $u \cdot (x \cdot v)^{\omega} \in L_n$ except that $x \cdot v = \emptyword$.
    Hence, $|\proEq^u_L| = 2$.
    It follows that the limit FDFA of $L_n$ has exactly $2 \times (n+1) + 1 + n+2 \in \Theta(n)$ states.

    Now we consider the index of $\proEq^u_R$ for a fixed $u \in \finwords$. 
    Similarly, when $q_{\bot} = \B(u)$, $|\proEq^u_R| = 1$ since for all $v \in \finwords$, we have $u \cdot x \cdot v \canoEq_{L_n} u \land u \cdot (x\cdot v)^{\omega} \notin L_n$ hold.
    Now we consider that $q_k = \B(u)$ with $0 \leq k \leq n$.
    Let $x_1, x_2 \in \finwords$.
    First, assume that $\B(u \cdot x_1 ) \neq \B(u \cdot x_2)$.
    W.l.o.g., let $q_j = \B(u \cdot x_2)$ with $0 \leq j \leq n$ and let $q_i = \B(u \cdot x_1)$ with either $i < j$ or $q_{i} = q_{\bot}$.
    We can easily construct a finite word $v$ such that $q_k = \B(u) =\B(u \cdot x_2 \cdot v)$, i.e., $u \cdot x_2 \cdot v \canoEq_{L_n} u$, and $u \cdot (x_2 \cdot v)^{\omega} \in L_n$.
    For example, we can let $v = (j+1) \cdots n \cdot 0 \cdots k$ if $j < k \leq n$.
    Hence, $u \cdot x_2 \cdot v \canoEq_{L_n} u \land u \cdot (x_2 \cdot v)^{\omega} \in L_n$ holds.
    On the contrary, it is easy to see that $q_{\bot} = \B(u \cdot x_1 \cdot v) = \trans(q_i, j+1)$ since either $j +1 > i + 1$ or $q_i = q_{\bot}$.
    In other words, we have $u \cdot x_1 \cdot v \not\canoEq_{L_n} u \land u \cdot (x_1 \cdot v)^{\omega} \notin L_n$ holds.
    By definition of $\proEq^u_R$, $x_1 \not\proEq^u_R x_2$.
    Hence, $|\proEq^u_R| \geq n + 2$.
    Next, we assume that $\B(u \cdot x_1) = \B(u \cdot x_2)$.
    For a word $v \in \finwords$, it is easy to see that $u \cdot x_1 \cdot v \canoEq_{L_n} u \Longleftrightarrow u \cdot x_2 \cdot v \canoEq_{L_n} u$.
    Moreover, since $u \cdot x_1 \cdot v \canoEq_{L_n} u$ implies $u \cdot (x_1 \cdot v)^{\omega} \in L_n$, we thus have that $u \cdot x_1 \cdot v \canoEq_{L_n} u \land u \cdot (x_1 \cdot v)^{\omega} \in L_n \Longleftrightarrow u \cdot x_2 \cdot v \canoEq_{L_n} u \land u \cdot (x_2 \cdot v)^{\omega} \in L_n $.
    In other words, $x_1 \proEq^u_R x_2$, which implies that $|\proEq^u_R| \leq n + 2$.
    Hence $|\proEq^u_R| = n + 2$ when $\B(u) \neq q_{\bot}$.
    It follows that the recurrent FDFA of $L_n$ has exactly $(n+2) \times (n + 1) + 1 + (n+2) \in \Theta(n^2)$ states.
    
    For the syntactic FDFA, since $\proEq^u_S$ refines $\proEq^u_R$~\cite{DBLP:journals/tcs/AngluinF16}, then $|\proEq^u_S| \geq |\proEq^u_R|$ for all $u \in \finwords$.
    The upper bound is proved similarly as for recurrent FDFAs.
    Therefore, the syntactic FDFA of $L_n$ also has $\Theta(n^2)$ states.

This completes the proof of the lemma.
\end{proof}

\section{Translations from FDFAs to NBAs}
\label{app:fdfa-to-nba}
It is possible to transform a canonical FDFA $\F$ of $L$ to an equivalent NBA $\A$~\cite{DBLP:conf/mfps/CalbrixNP93,AngluinBF18,DBLP:journals/iandc/LiCZL21}.

In the following, we only briefly describe how we construct a NBA from an FDFA.
Angluin and Fisman proved in~\cite{AngluinBF18} that every saturated FDFA $\F$ can be polynomially translated to an equivalent NBA $\A[\F]$.
In fact, the requirement for $\F$ being saturated is somewhat strong;
we only need $\F$ to be almost saturated.

The translation given in~\cite{DBLP:conf/mfps/CalbrixNP93,AngluinBF18,DBLP:journals/iandc/LiCZL21} works as follows.
Let $\F = (\M, \setnocond{\N^q})$ be an almost saturated FDFA, where $\M = (\alphabet, Q, \iota, \trans)$, and for each state $q \in Q$, there is a progress DFA $\N^q = (\alphabet, Q_q, \iota_q, \trans_q, F_q)$.
Recall that $(A)^s_f$ denotes the DFA $A$ where $s$ is the initial state and $f$ is the sole final state.
By Definition~\ref{def:acc-fdfa}, we have that $\upword{\F} = \setnocond{\alpha \in \infwords: \alpha \emph{ is accepted by }\F}$, where $\alpha$ is accepted if there is a decomposition $(u, v)$ of $\alpha$, such that $\M(u) = \M(uv)$, and $\N^q(v) \in F_q$ where $q = \M(u)$.
This implies that a word $\alpha\in \upword{\F}$ can be decomposed into two parts $u$ and $v$, such that $u$ is accepted by the DFA $\M^{\iota}_q$ and $v$ by the DFA $(\N^q)^{\iota_q}_{f}$ where $f = \N^q(v)$.
Hence, $\upword{\F} = \bigcup_{q\in Q, f \in F_q} \finlang{M^{\iota}_q} \cdot N_{(q, f)}$, where
$N_{(q, f)} = \setnocond{v^{\omega} \in \infwords: v \in \poswords, q = \M^q_q(v), v \in \finlang{(\N^q)^{\iota_q}_q}}$ is the set of all infinite repetitions of the finite words $v$ accepted by $(\N^{q})^{\iota_q}_f$.

It is hard to construct a NBA to accept exactly $N_{(q, f)}$.
However, it suffices to under approximate $N_{(q, f)}$ with the DFA $P_{(q, f)} = \M^q_q \times (\N^q)^{\iota_q}_q \times (\N^q)^f_f$, where $\times$ stands for the intersection product between DFAs.
On one hand, the DFA $\M^q_q \times (\N^q)^{\iota_q}_q$ makes sure that for a word $v \in \finlang{\M^q_q \times (\N^q)^{\iota_q}_q}$ and $u \in \finlang{\M^{\iota}_q}$, it follows that $q = \M(u) = \M(uv)$.
On the other hand, $(\N^q)^f_f$ ensures that $v, v^k \in \finlang{(\N^q)^{\iota_q}_f}$ for all $k \geq 1$.
One can construct a NBA $\A[\F] = \bigcup_{q \in Q, f \in F_q} \finlang{\M^{\iota}_q} \cdot P^{\omega}_{(q,f)}$ to under approximate $\upword{\F}$~\cite{DBLP:journals/iandc/LiCZL21}.

It is worth noting that we can construct easily a DBA that accepts $P^{\omega}_{(q,f)}$ from the DFA $P_{(q,f)}$ by redirecting all incoming transitions of final states to the initial state and mark them as $\accd$-transitions.
This way, we obtain a LDBA $\sd[\F]$ that recognizes $\upword{\F}$, which allows easier determinization algorithm~\cite{DBLP:conf/tacas/EsparzaKRS17,DBLP:conf/cav/LiTFVZ22}.
This construction of LDBAs is much easier than the one proposed in~\cite{DBLP:journals/iandc/LiCZL21} where the acceptance condition is defined on states, rather than transitions.

Since the four types of canonical FDFAs are all saturated, Corollary~\ref{coro:almost-saturated} immediately follows.
\begin{corollary}\label{coro:almost-saturated}
    Let $L$ be an $\omega$-regular language.
    Then its periodic, syntactic, recurrent and limit FDFAs are almost saturated.
\end{corollary}

Let $n$ is the number of states in the leading DFA $\M$ and $k$ is the largest number of states of progress DFAs of $\F$.
For each pair $q \in \states, f \in F_q$, the constructed NBA/DBA accepting $P_{(q,f)}$ has $nk^2$ states, and there are at most $nk$ such pairs;
So, all four types of canonical FDFAs can be polynomial translated to equivalent NBA/LDBAs with $\bigO(n^2 k^3)$ states. 

For the variant limit FDFA $\F_B$, there is at most one final state in each progress DFA.
So, the equivalent NBA for $\F_B$ has $\bigO(n^2 k^2)$ states.

\section{Proof of Lemma~\ref{lem:dba-lang-co-safety}}
\label{app:dba-cosafety}
\dbaCoSafe*
\begin{proof}
    The proof is inspired and adapted from the proof of \cite[Lemma~10]{DBLP:conf/icalp/BohnL22}.
    
    We let $\D = (\T, \accd)$ be a DBA of $L$, where $\T =( \states, \alphabet, q_0, \trans)$ is the TS of $\D$ and $\accd$ is the set of accepting transitions.
    We assume that $\D$ is complete in the sense that for every state $q \in \states$ and $\sigma \in \alphabet$, we have that $\trans(q, \sigma) \in \states$. 
    
    For two different states $q_1, q_2 \in \states$, we define an equivalence relation $\canoEq_{\D}$ where $q_1 \canoEq_{\D} q_2$ if and only if $\lang{\D^{q_1}} = \lang{\D^{q_2}}$ where $\D^q$ is the DBA obtained from $\D$ by setting the initial state to $q \in \states$.
    Let $U_q = \setnocond{ u \in \finwords: \trans(q_0, u) = q}$.
    Let $U_{[q]_{\canoEq_{\D}}} = \cup_{p \in [q]_{\canoEq_{\D}}} U_p$ where $[q]_{\canoEq_{\D}}$ is the equivalence class of $\canoEq_{\D}$ that $q$ belongs to.
    Clearly, $U_{[q]_{\canoEq_{\D}}}$ is an equivalence class $\class{u}$ of $\canoEq$ defined with respect to $L$ where $u\in U_{[q]_{\canoEq_{\D}}}$.

    Now consider the periodic finite words for each state $q \in \states$.
    Let $V_q = \setnocond{x \in \poswords: \forall v \in \finwords.\ \text{ if } q \xrightarrow{x \cdot v} q.\ (x\cdot v)^{\omega} \in \lang{\D^q}}$.
    That is, a word $x$ belongs to $V_q$ iff for every $v \in \finwords$, if $\D$ takes a round trip from $q$ back to itself over $x\cdot v$, the run must go through a $\accd$-transition.
    We first prove that $V_q$ is regular.
    We can construct the DFA $D_q $ of $V_q$ from the TS $\T$ by first removing all $\accd$-transitions in $\T$, resulting a TS $\T'$, and then collect all the transitions $(p, \sigma, q)$ in a set $\beta$ such that $p$ and $q$ are in the different SCCs of the reduced TS $\T'$.
    We then define $D_q = (\states \cup \setnocond{\top}, \alphabet, q, \trans_D, F = \setnocond{\top})$ where
    (1) for a state $p \in \states$, $\sigma \in \alphabet$ and $q = \delta(p, \sigma)$, $\trans_D(p, \sigma) = q$ if $(p, \sigma, q) \notin \accd \cup \beta$ and otherwise $\trans_D(p, \sigma) = \top$;
    and (2) $\trans_D(\top, \sigma) = \top$ for all $\sigma \in \alphabet$.
    
    Next we prove that $\finlang{D_q} = V_q$.
    First, let $x \in \finlang{D_q}$ and we want to prove that $x \in V_q$.
    Obviously, the last transition of $\D$ over $x$ from $q$ will be either a $\accd$-transition or a transition jumping between two SCCs in the reduced $\T'$.
    If it is a $\accd$-transition, obviously, we have that for all $v \in \finwords$, if $q \xrightarrow{x\cdot v} q$, then it must visit a $\accd$-transition. 
    Hence, $(xv)^{\omega} \in \lang{\D^q}$.
    If it is a transition jumping between different SCCs, it would be the case that either $\D$ does not go back to $q$ over $xv$ or it must be visiting a $\accd$-transition, since in the reduced TS $\T'$, they can not reach each other.
    Therefore, $x \in V_q$.
    Now let $x \in V_q$ and we want to prove that $x \in \finlang{D_q}$.
    Let $p = \trans(q, x)$ in $\D$.
    If $p$ and $q$ lie in two different SCCs of $\D$, then it is impossible to find a $v \in \finwords$ such that $ p \xrightarrow{ v} q$, otherwise, $p$ and $q$ will belong to the same SCC of $\D$.
    In this case, there will be a transition between different SCCs along the way from $q$ to $p$ over $xv$, which of courses also separates these two SCCs in the reduced TS $\T'$. 
    Thus, there will be a prefix of $x$ accepted by $D_q$, so $x$ is also accepted by $D_q$ as $\top$ is a sink final state.
    Now assume that $p$ and $q$ are in the same SCC of $\D$.
    At state $p$, for each $v \in \finwords$ such that $q \xrightarrow{x} p \xrightarrow{v} q$, we have that $(x \cdot v)^{\omega} \in \lang{\D^q}$.
    There must be some $\accd$-transition visited along the way from $q$ back to itself.
    It follows that in the reduced TS $\T'$, it is impossible to reach $p$ from $q$.
    In other words, $q$ and $p$ are not in the same SCC of $\T'$.
    So, the run from $q$ to $p$ over $x$ must visit some transition jumping between two different SCCs.
    Again, this means that there will be a prefix of $x$ accepted by $D_q$.
    So $x$ will also be accepted by $D_q$.
    Therefore, $V_q$ is a regular language.

    Now, for an equivalence class $[q]_{\canoEq_{\D}}$, we define $V_{[q]_{\canoEq_{\D}}} = \bigcap_{p \in [q]_{\canoEq_{\D}}} V_p$.
    So, $V_{[q]_{\canoEq_{\D}}}$ is also a regular language.
    Let $u$ be a word in $U_{[q]_{\canoEq_{\D}}}$.

    Let $V_u = \setnocond{ x \in \poswords: \forall v \in \finwords.\ u \cdot (x \cdot v) \canoEq u \implies u\cdot (x\cdot v)^{\omega} \in L} $.
    Next, we prove that $V_{u} \equiv V_{[q]_{\canoEq_{\D}}}$.
    Let $p = \trans(\init, u)$.
    
    Let $x \in V_{[q]_{\canoEq_{\D}}}$ and we want to prove that $x \in V_u$.
    That is, we need to prove that for all $v \in \finwords$, we have that $u \cdot (x \cdot v) \canoEq u \implies u \cdot (x \cdot v)^{\omega} \in L$.
    First, if $u \cdot (x \cdot v) \not \canoEq u$, then $x \in V_u$ holds trivially.
    Otherwise we have that $u \cdot x \cdot v \canoEq u$, which implies that $\trans(q_0,u \cdot (x \cdot v)^k) \canoEq_{\D} \trans(q_0, u)$ for all $k \geq 0$.
    Thus, we will have a run $\rho = q_0 \xrightarrow{u} q_1 \xrightarrow{x \cdot v} \cdots$ of $\D$ over $u \cdot (xv)^{\omega}$ where $q_i \in [q]_{\canoEq_{\D}}$ for all $i > 0$.
    There must be some state $q$ occurs for an inifinite set of indices $I = \setnocond{i \in \naturals: q = q_i}$.
    For each $q_i \in [q]_{\canoEq_{\D}}$, we have that $x \in V_{q_i}$.
    First, $x \in V_{p}$ for all states $p \in [q]_{\canoEq_{\D}}$, so for every two pairs of integers $i, j \in I$ with $i < j$, there must be a $\accd$-transition along the way from $q_i$ to $q_j$. 
    It follows that $u \cdot (x \cdot v)^{\omega} \in \lang{\D^q}$ holds.
    Hence, $x \in V_u$ holds as well, since $ u \cdot x \cdot v \canoEq u \implies u \cdot (x \cdot v)^{\omega} \in L$ holds for all $v \in \finwords$.

    Now assume that $x \notin V_{[q]_{\canoEq_{\D}}}$ and we want to prove that $x \notin V_u$ holds.
    Assume by contradiction that $x \in V_u$.
    Since $x$ does not belong to $V_{[q]_{\canoEq_{\D}}}$, then there exists a state $r \in [q]_{\canoEq_{\D}}$ such that $x \notin V_r$.
    That is, there exists a word $v\in \finwords$ such that $r \xrightarrow{x\cdot v} r$ and $(x \cdot v)^{\omega} \notin \lang{\D^r}$.
    Since $p \canoEq_{\D} r$, i.e., $\lang{\D^p} = \lang{\D^r}$, $(x \cdot v)^{\omega} \notin \lang{\D^p}$ as well.
    It then follows that $u \cdot (x \cdot v) \canoEq u$ and $u \cdot (x\cdot v)^{\omega} \notin L$, which contradicts that $x \in V_u$.
    Therefore, $x \notin V_u$.

    Hence, $V_u = V_{[q]_{\canoEq_{\D}}}$.
    Now we show that $V_u$ is an equivalence class of $\proEq^u_L$ as follows.
    On one hand, for every two different words $x_1, x_2 \in V_u$, we have that $x_1 \proEq^u_L x_2$, which is obvious by the definition of $V_u$.
    On the other hand, it is easy to see that $x' \not\proEq^u_L x$ for all $x' \notin V_u$ and $x \in V_u$ because there will exists some $v \in \finwords$ such that $u \cdot x' \cdot v \canoEq u$ but $u\cdot (x' \cdot v)^{\omega} \notin L$.
    Hence, $V_u$ is indeed an equivalence class of $\proEq^u_L$.
    Obviously, $V_u \subseteq \finlang{\N^u}$, as we can let $v = \emptyword$, so for every word $x \in V_u$, we have that $u \cdot x \canoEq u \implies u \cdot x^{\omega} \in L$.
    Let $\rep{x} = \N^u(x)$ for a word $x \in V_u$.
    It follows that $\rep{x}$ is a final state of $\N^u$ and we have $[\rep{x}]_{\proEq^u_L} = V_u$.
    Thus, we complete the proof of the lemma.

\end{proof}

\section{Proof of Theorem \ref{thm:last-fdfa-rep}}
\label{app:proofofT6}
\thmLastFdfaRep*
\begin{proof}
The first claim follows from the restriction to finite indices in the definition (we have seen that they exist, and that we can, e.g., choose limit RC).


    To show $\upword{\F_L} \subseteq \upword{L}$,
    assume that $w \in \upword{\F_L}$.
    By Definition~\ref{def:acc-fdfa}, a UP-word $w$ is accepted by $\F_L$ if there exists a decomposition $(u, v)$ of $w$ such that $\M(u) = \M(u \cdot v)$ (equivalently, $u \cdot v \canoEq u$) and $v \in \finlang{\N^{\rep{u}}}$ where $\rep{u} = \M(u)$.
    Here $\rep{u}$ is the representative word for the equivalence class $[u]_{\canoEq}$.
    Similarly, let $\rep{v} = \N^{\rep{u}}(v)$.
    By Definition \ref{def:nondet-proeq}, we have $\rep{u} \cdot \rep{v} \canoEq \rep{u}\implies \rep{u} \cdot \rep{v}^{\omega} \in L$ holds as $\rep{v}$ is a final state of $\N^{\rep{u}}$.
    Since $v \proEq_{\rep{u}} \rep{v}$ (i.e., $\N^{\rep{u}}(v) = \N^{\rep{u}}(\rep{v})$), $\rep{u} \cdot v \canoEq \rep{u}\implies \rep{u} \cdot v^{\omega} \in L$ holds as well.
    It follows that $u \cdot v \canoEq u \implies u \cdot v^{\omega} \in L$ since $u \canoEq \rep{u}$ and $u \cdot v \canoEq \rep{u} \cdot v$ (equivalently, $\M(u \cdot v) = \M(\rep{u} \cdot v)$).
    Together with the assumption that $\M(u \cdot v) = \M(u)$ (i.e, $u \canoEq u \cdot v$), we then have that $u \cdot v^{\omega} \in L$ holds.
    So, $\upword{\F_L} \subseteq \upword{L}$ also holds.
    
    To show that $\upword{L} \subseteq \upword{\F_L}$ holds, let $w \in \upword{L}$.
    For a UP-word $w \in L$, we can find a normalized decomposition $(u, v)$ of $w$ such that $w = u\cdot v^{\omega}$ and $u \cdot v \canoEq u$ (i.e., $\M(u) = \M(u \cdot v)$), since the index of $\canoEq$ is finite (cf.~\cite{DBLP:journals/tcs/AngluinF16} for more details).
    Let $\rep{u} = \M(u)$ and $\rep{v} = \N^{\rep{u}}(v)$.
    Our goal is to prove that $\rep{v}$ is a final state of $\N^{\rep{u}}$.
    Since $u \canoEq \rep{u}$ and $u\cdot v^{\omega} \in L$, then $\rep{u}\cdot v^{\omega} \in L$ holds.
    Moreover, $\rep{u} \cdot v \canoEq \rep{u}$ holds as well because $\rep{u} = \M(\rep{u}) = \M(u)= \M(\rep{u} \cdot v) = \M(u \cdot v)$.
    (Recall that $\M$ is deterministic.)
    We now have that $v \in C_u$, so that $C_{\rep{u}} \cap \Sigma^*/_{\proEq^u_N}$ is good (as $u\cdot v^\omega \in L$).
    We also have that $\rep{v} \proEq^u_N v$, so that $[\rep{v}]_{\proEq^u_N}$ is accepting.
    Hence, $\rep{v}$ is a final state, and $(u, v)$ therefore accepted by $\F_L$, i.e., $w \in \upword{\F_L}$.
    It follows that $\upword{L} \subseteq \upword{\F_L}$.

    Now we prove that $\F_L$ is saturated.
    Let $w$ be a UP-word.
    Let $(u, v)$ and $(x, y)$ be two normalized decompositions of $w$ with respect to $\M$ (or, equivalently, to $\canoEq$).
    We have seen that $(u, v)$ is accepted by $\F_L$ iff $u \cdot v^\omega = x \cdot y^\omega \in \upword{L}$, which is the case iff $(x, y)$ is accepted by $\F_L$ with the same argument.
\end{proof}

\section{Active learning of limit FDFAs}
\label{app:learning}

First, there are two roles, namely the learner and an oracle in the active learning framework~\cite{DBLP:journals/iandc/Angluin87}.
The task of the learner is to learn an automaton representation of an unknown language $L$ from the oracle.
The learner can ask two types of queries about $L$, which will be answered by the oracle.
A membership query is about whether a word $w$ is in $L$;
an equivalence query is to ask whether a given automaton recognizes the language $L$.
If the oracle returns positive answer to equivalence query, then the learner has completed the task and output the correct automaton;
otherwise, the learner will receive a counterexample which will then be used to refine current hypothesis.

Angluin and Fisman proposed a learning framework in \cite{DBLP:journals/tcs/AngluinF16} to learn the classical three types of FDFAs.
We show that our limit FDFA can easily fit into this learning framework.
The learner $L^{\omega}$ is described in the following framework.
We refer to \cite{DBLP:journals/tcs/AngluinF16} for details about the learning framework.
We mainly use the notations and description from \cite{DBLP:journals/tcs/AngluinF16} in the following.
As usual, the framework makes use of the notion of \emph{observation tables}.
An observation table is a tuple $\T = (S, \rep{S}, E, T)$ where $S$ is a prefix-closed set of finite words, $E$ is a set of experiments trying to distinguish the strings in $S$, and $T: S \times E \rightarrow D$ stores
the element (membership query results) in entry $T(s, e)$ an element in some domain $D$, where $s \in S$ and $e \in E$.
For our limit FDFA, $D$ is purely a Boolean values $\setnocond{\top, \bot}$.
We usually determine when two strings $s_1, s_2 \in S$
should be considered not equivalent depending on the RC we are using. 
The component $\rep{S} \subseteq S$ is the subset considered as representatives of the equivalence classes, i.e., the state names of the constructed DFA.
A table is said to be \emph{closed} if $S$ is prefix closed and for every $s \in \rep{S}$ and $\sigma \in \alphabet$, we
have $s\sigma \in S$.
The procedure \emph{CloseTable} uses two sub-procedures $\fn{ENT}$ and $\fn{DFR}$ to make a given observation closed.
Here $\fn{ENT}$ is used to fill in the
entries of the table by means of asking membership queries.
The procedure $\fn{DFR}$ is used to determine which row (words) of the table should be distinguished.
A learning procedure usually begins with create an initial observation table by asking membership queries, close the table with $\fn{ENT}$ and $\fn{DFR}$ procedures, and then construct an hypothesis automaton for asking equivalence query.
The learner should be able to use the counterexample to the equivalence query to find new experiments for discovering new equivalence classes.

We now give the subprocedures for learning our limit FDFAs.

\begin{algorithm}
\caption{The learner $L^{\omega}$ in \cite{DBLP:journals/tcs/AngluinF16}}\label{alg:cap}
Initialize leading table $\T = (S, \Tilde{S}, E, T)$ with $S = \Tilde{S} = \setnocond{\emptyword}, E = \setnocond{(\emptyword, \sigma): \sigma \in \alphabet}$\;
$\tfn{CloseTable}(\T, \fn{ENT}_1, \fn{DFR}_1)$ and let $\M = \fn{Aut}_1(\T)$\;
\ForAll{ $u \in \Tilde{S}$}
{
Initialize $\T_u = (S_u, \Tilde{S}_u, E_u, T_u)$, with $S_u = \Tilde{S}_u = E_u = \setnocond{\emptyword}$\;
$\tfn{CloseTable}(\T_u, \fn{ENT}^u_2, \fn{DFR}^u_2)$ and let $\A_u = \fn{Aut}_2(\T_u)$\;
}

\While{ $\text{true}$}
{
  Let $(a, u, v)$ be the oracle's response for equivalence query $\mathcal{H} = (\M, \setnocond{\A_u})$\;
  \If{ $a = $ ``yes''}{
    break\;
  }
  Let $(x, y)$ be the normalized decomposition of $(u, v)$ w.r.t $\M$\;
  Let $\Tilde{x} = \M(x)$\;
  \If{$\fn{MQ}(x, y ) \neq \fn{MQ}(\Tilde{x}, y)$}{
    $E = E \cup \tfn{FindDistinguishingExperiment}(x,y)$\;
    $\tfn{CloseTable}(\T, \fn{ENT}_1, \fn{DFR}_1)$ and let $\M = \fn{Aut}_1(\T)$\;
  }
  \Else{
    $E_{\Tilde{x}} = E_{\Tilde{x}} \cup \mathit{FindDistinguishingExperiment}(\Tilde{x},y)$\;
    $\tfn{CloseTable}(\T_{\Tilde{x}}, \fn{ENT}^{\Tilde{x}}_2, \fn{DFR}^{\Tilde{x}}_2)$ and let $\A_{\Tilde{x}} = \mathit{Aut}_2(\T_{\Tilde{x}})$\;
  }
}
\end{algorithm}

We let $\fn{MQ}(x, y)$ be the result of the membership query $\omega$-word $x\cdot y^{\omega}$ to the oracle.
The procedures $\fn{ENT}_1$ and $\fn{DFR}_1$ and $\fn{Aut}_1$ are the same for all four types of FDFAs.
More precisely, for $u, x, y \in \finwords$, $\fn{ENT}_1(u,(x, y))  = \fn{MQ}(u \cdot x, y)$;
for two finite row words $u_1, u_2 \in S$, $\fn{DFR}_1(u_1, u_2)= \top$ iff there exists $(x, y) \in E$ such that $T(u_1, (x,y)) \neq T(u_2, (x, y))$.
That is, we can use $x\cdot y^{\omega}$ to distinguish the finite words $u_1$ and $u_2$ according to $\canoEq$.
The procedure $\fn{Aut}_1$ is simply to construct the leading DFA without final states from $\T$, by Definition~\ref{def:dba-construction}.
When learning our limit FDFAs, for $u, x, v \in \finwords$, we define $\fn{ENT}^u_2(x,v) = \top$ if $\M(u x\cdot v) \neq \M(u)$ or $\fn{MQ}(u, x \cdot v) = \top$ holds, corresponding to whether $ux\cdot v\canoEq u \implies u \cdot (xv)^{\omega} \in L$ holds in Definition~\ref{def:limit-rcs};
for two finite row words, $x_1, x_2 \in S_u$, $\fn{DFR}^u_2(x_1, x_2)$ returns true if there exists $v\in E$ such that $T_u(x_1,v) \neq T_u(x_2, v)$.
The procedure $\fn{Aut}_{u}(\T_u)$ not only constructs the TS but also set a state $x$ as accepting if $T_u(x, \emptyword) = \top$.
Note that here $T_u(x, v)$ stores the result of whether $(\M(u \cdot xv) = \M(u)) \implies \fn{MQ}(u, xv)$.

To be consistent with the notations in \cite{DBLP:journals/tcs/AngluinF16}, we also denote by $\subword{\run}{i}{k}$ the subsequence of $\run$ starting at the $i$-th element and ending at the $k$-th element (inclusively) when $i \leq k$, and the empty sequence $\emptyword$ when $i > k$.
However, the first element will be $\run[1]$ instead of $\run[0]$ in the main content.

Now we provide more details in learning our limit FDFAs and also prove that the learner $L^{\omega}$ will make progress in every iteration.
We assume that now we have received the counterexample $(u,v)$ in the algorithm to current hypothesis and we prove that our limit FDFA learner is able to make use of $(u,v)$ to refine current FDFA.

Let $(x, y)$ be the normalized decomposition of the counterexample $u\cdot v^{\omega}$ with respect to $\M$ and let $\rep{x} = \M(x)$.
If $\fn{MQ}(x,y) \neq \fn{MQ}(\rep{x},y)$, then we know that $x\not\canoEq \rep{x}$.
So, we can find an experiment as follows:
let $n = |x|$ and for $1 \leq i \leq n$, let $s_i = \M(\wordletter{x}{1\cdots i})$ be state/word that $\M$ arrives after reading the first $i$ letters of $x$.
Recall that $s_i$ is also the representative word of $ \M(\wordletter{x}{1\cdots i})$.
In particular, $s_0 = \M(\emptyword) = \emptyword$ and $s_n = \M(x) = \rep{x}$.
Thus, we can construct the sequence, $\fn{MQ}(s_0 \cdot \wordletter{x}{1\cdots n}, y), \fn{MQ}(s_1 \cdot \wordletter{x}{2\cdots n}, y), \fn{MQ}(s_2 \cdot \wordletter{x}{3\cdots n}, y),\cdots, \fn{MQ}(s_n \cdot \wordletter{x}{n+1\cdots n}, y)$.
Obviously, this sequence has different results for the first and last elements since $\fn{MQ}(s_0 \cdot \wordletter{x}{1\cdots n}, y) \neq \fn{MQ}(s_n, y)$, where $s_n = \rep{x}$.

Therefore, there must exist the smallest $j\in [1\cdots n]$ such that $\fn{MQ}(s_{j-1} \cdot \wordletter{x}{j\cdots n}, y) \neq \fn{MQ}(s_j \cdot \wordletter{x}{j+1\cdots n}, y)$,
It follows that we can use the experiment $e = (u[j + 1 \cdots n], v)$ to distinguish $s_{j-1} \cdot \wordletter{x}{j}$ and $s_{j}$.

Otherwise if $\fn{MQ}(x,y) = \fn{MQ}(\rep{x},y)$, we need to similarly refine current $\A_{\rep{x}}$.
Similarly, we let $n = |y|$ and $s_i = \A_{\rep{x}}(y[1\cdots i])$.
We also consider a sequence $(m_0, c_0), \cdots, (m_n, c_n)$ where $m_i = \top$ iff $ \rep{x} = \M(\rep{x} \cdot s_i \cdot \wordletter{y}{i+1 \cdots n}) $ and $c_i = \top$ iff $\rep{x} \cdot (s_i \cdot \wordletter{y}{i+1 \cdots n})^{\omega} \in L$.
First, we know that $m_0 = \top $ and $m_n = \top $ since $(x, y)$ is a normalized decomposition of $u\cdot v^{\omega}$, i.e., $\rep{x} = \M(x) = \M(x\cdot y) = \M(\rep{x}\cdot y)$.
Since $(x, y)$ is a counterexample to current hypothesis $\mathcal{H}$, we know that either the normalized decomposition $(x, y)$ is not accepted by $\mathcal{H}$ and $xy^{\omega} \in L$ or $(x, y)$ is accepted by $\mathcal{H}$ and $xy^{\omega} \notin L$.
Therefore, one out of $(m_0, c_0)$ and $(m_n, c_n)$ must be $(\top, \top)$ and the other is not.
That is, either $m_0 \implies c_0$ or $m_n \implies c_0$ holds.
There must be the smallest $j \in [1\cdots n]$ such that $m_{j-1} \implies c_{j-1}$ and $m_j \implies c_j$ differs.
W.l.o.g., we let $m_{j-1} \implies c_{j-1}$ hold.
In this case, we can set the experiment $e = \wordletter{y}{j+1 \cdots n}$ to distinguish $s_{j-1} \cdot \wordletter{y}{j}$ and $s_j$ since we have $\rep{x} = \M(\rep{x} \cdot s_{j-1} \cdot \wordletter{y}{j\cdots n}) \implies \rep{x} \cdot (s_{j-1} \cdot \wordletter{y}{j\cdot n})^{\omega} \in L$ but $\rep{x} = \M(\rep{x} \cdot s_{j} \cdot \wordletter{y}{j+1\cdots n}) \implies \rep{x} \cdot (s_{j} \cdot \wordletter{y}{j+1\cdots n})^{\omega} \in L$ does not hold.

We can see that every time we received a counterexample from the oracle, either the leading DFA $\M$ or the progress DFA $\A_{\rep{x}}$ will add at least state.
Since the limit FDFA $\F_L$ has finite number of states, $\mathcal{H}$ will eventually be $\F_L$ in the worst case. 

\begin{corollary}
    The limit FDFAs can be learned with membership and equivalence queries in time in polynomial in the size of canonical limit FDFAs.
\end{corollary}


\end{document}